\documentclass[11pt]{article}
\usepackage{graphicx}
\usepackage[utf8]{inputenc} 

\usepackage[export]{adjustbox}
\pdfoutput=1

\usepackage{amsmath,amsfonts,amssymb,epsfig}
\usepackage{slashed}
\numberwithin{equation}{section}

\usepackage{graphicx}
\usepackage{cancel}
\usepackage{cite}
\usepackage{color}

\usepackage{xspace}

\newcommand{\SARAH}{{\tt SARAH}\xspace}

\newcommand{\exclude}[1]{}
\def\nn{\nonumber}

\def\beq{\begin{equation}}
\def\eeq{\end{equation}}
\def\bal{\begin{align}}
\def\eal{\end{align}}

\def\lagr{{\cal L}}

\def\s2b{s_{2\beta}}
\def\c2b{c_{2\beta}}

\long\def\symbolfootnote[#1]#2{\begingroup%
\def\thefootnote{\fnsymbol{footnote}}\footnote[#1]{#2}\endgroup}

\def\2b2[#1,#2][#3,#4]{\left( \begin{array}{cc} #1 & #2 \\ #3 & #4 \end{array}
\right)}
\def\3b3[#1,#2,#3][#4,#5,#6][#7,#8,#9]{\left( \begin{array}{ccc} #1 & #2 &#3 \\
#4 & #5 & #6\\#7&#8&#9\end{array} \right)}
\def\thv[#1,#2,#3]{\left( \begin{array}{c} #1 \\ #2 \\ #3 \end{array} \right)}
\def\twv[#1,#2]{\left( \begin{array}{c} #1 \\ #2 \end{array} \right)}

\def\twomat[#1,#2][#3,#4]{\left( \begin{array}{cc} #1 & #2 \\ #3 & #4 \end{array} \right)}
\def\threemat[#1,#2,#3][#4,#5,#6][#7,#8,#9]{\left( \begin{array}{ccc} #1 & #2 & #3\\ #4 & #5 & #6 \\ #7 & #8 & #9 \end{array} \right)}
\def\twovec[#1,#2]{\left( \begin{array}{c} #1  \\ #2 \end{array} \right)}

\def\ov{\overline}

\def\msbar{{\ov {\rm MS}}}
\def\beq{\begin{equation}}
\def\eeq{\end{equation}}
\def\bea{\begin{eqnarray}}
\def\eea{\end{eqnarray}}

\def\Rt{\tilde{R}}

\def\Otl{\mathcal{O}({\rm 2\ loop})}

\def\sq2{\sqrt{2}}
\def\drbar{{\ensuremath{ \overline{\rm DR}^\prime}}\xspace}
\def\msbar{{\ensuremath{\overline{\rm MS}}}\xspace}
\def\dimreg{\mathrm{dim.\ reg}}

\newcommand{\vtree}{V^{(0)}}

\newcommand{\veff}{V_{\text{eff}}}
\newcommand{\vone}{V^{(1)}}
\newcommand{\vtwo}{V^{(2)}}
\newcommand{\vtwos}{V^{(2)}_S}

\newcommand{\vtwofs}{V^{(2)}_{FS}}
\newcommand{\veffr}{\hat{V}_{\text{eff}}}
\newcommand{\gbm}{m_G^2}

\newcommand{\vtwor}{\hat{V}^{(2)}}

\DeclareMathOperator{\llog}{\overline{\text{log}}}
\DeclareMathOperator{\blog}{\overline{\text{log}}}

\newcommand{\propS}{{\rm S}}
\newcommand{\propU}{{\rm U}}
\newcommand{\propM}{{\rm M}}
\newcommand{\propV}{{\rm V}}
\newcommand{\propW}{{\rm W}}
\newcommand{\propX}{{\rm X}}
\newcommand{\propY}{{\rm Y}}
\newcommand{\propZ}{{\rm Z}}
\newcommand{\Fbar}{\overline{F}}
\newcommand{\re}{\mathrm{Re}}

\begin{document}

\begin{titlepage}

\begin{flushright}
\end{flushright}
\begin{center}

\vspace{1cm}

{\LARGE \bf Avoiding the Goldstone Boson Catastrophe} 
\vskip 0.3cm
{\LARGE \bf in general renormalisable field theories at two loops}

\vspace{1cm}

{\Large Johannes~Braathen,$^{\!\!\!\,a,b}$\symbolfootnote[1]{{\tt braathen@lpthe.jussieu.fr}}~ 
and Mark~D.~Goodsell$^{\,a,b}$\symbolfootnote[2]{{\tt goodsell@lpthe.jussieu.fr}}}

\vspace*{5mm}

{\sl ${}^a$ LPTHE, UPMC Univ.~Paris 06,
  Sorbonne Universit\'es, 4 Place Jussieu, F-75252 Paris, France}
\vspace*{2mm}\\
{\sl ${}^b$ LPTHE, CNRS, 4 Place Jussieu, F-75252 Paris, France }
\end{center}

\vspace{0.7cm}

\abstract{We show how the infra-red divergences associated to Goldstone bosons in the minimum condition of the two-loop Landau-gauge effective potential can be avoided in general field theories. This extends the resummation formalism recently developed for the Standard Model and the MSSM, and we give compact, infra-red finite expressions in closed form for the tadpole equations. We also show that the results at this loop order are equivalent to (and are most easily obtained by) imposing an ``on-shell'' condition for the Goldstone bosons. Moreover, we extend the approach to show how the infra-red divergences in the calculation of the masses of neutral scalars (such as the Higgs boson) can be eliminated. For the mass computation, we specialise to the gaugeless limit and extend the effective potential computation to allow the masses to be determined without needing to solve differential equations for the loop functions -- opening the door to fast, infra-red safe determinations of the Higgs mass in general theories. 
}

\vfill

\end{titlepage}

\tableofcontents

\setcounter{footnote}{0}

\section{Introduction}
\label{SEC:intro}

The discovery of the Higgs boson has added a wealth of electroweak precision observables, chief among them being its mass, which is remarkably known to within a  few hundred MeV. The combined data can be used to determine the fundamental lagrangian parameters of the theory, such as the Higgs mass-squared parameter and quartic coupling, and then make predictions for future measurements, such as the Higgs self-coupling; or to provide the starting point for extrapolations of the potential to high energy scales to study unification or vacuum stability. 

In the context of theories beyond the Standard Model, the Higgs mass-squared parameter and also its mass are very sensitive to new particles, and can thus be used to constrain new physics. However, precisely because of this sensitivity, the accuracy of the theoretical calculation is typically much poorer than the experimental measurement, in particular for supersymmetric field theories (which remain renormalisable up to high scales) and there has therefore been a significant effort to improve these calculations.

Typically, a subset of the scalar mass-squared parameters in the lagrangian are determined from the tadpole equations, which are the (first) derivatives of the effective potential. In the Standard Model the full effective potential was computed to two-loop order in \cite{Ford:1992pn}, with the 3-loop leading contributions involving the strong and Yukawa couplings found in \cite{Martin:2013gka},
and the 4-loop part at leading order in QCD in \cite{Martin:2015eia}. 
However, for general renormalisable theories the potential is only known to two-loop order\footnote{Although note that results in the zero momentum approximation are available for the Higgs boson \emph{mass} in  the Minimal Supersymmetric Standard Model (MSSM) up to partial three-loop order \cite{Harlander:2008ju,Kant:2010tf}.} via the expressions given in \cite{Martin:2001vx} which were used in \cite{Goodsell:2015ira} to derive the tadpole equations (while the diagrams for the masses were already given in \cite{Martin:2003it}). 

For reasons of calculational simplicity, the effective potential beyond one loop has been calculated only in the Landau gauge, which means that the would-be Goldstone boson is treated as an actual massless Goldstone boson. Unfortunately, this leads to a technical problem known as the Goldstone Boson Catastrophe: the mass-squared lagrangian parameter of the Goldstone boson determined from the tadpole equations\footnote{Note that we take the expectation values to be fixed and loop-correct the mass-squared terms rather than vice versa.} is small and can even be negative (as opposed to the pole mass, which is always zero) and this causes the loop integrals for the tadpoles to diverge or be complex. While this problem can in principle be circumvented by dropping the complex parts and changing the renormalisation scale to attempt to find  non-negligible positive mass-squareds, this is not easy to implement consistently.

A solution for the tadpoles was proposed in \cite{Martin:2014bca,Elias-Miro:2014pca} for the Standard Model and applied to the MSSM in \cite{Kumar:2016ltb} (see also \cite{Andreassen:2014eha,Espinosa:2016nld,Espinosa:2016uaw} for recent related work): (a subset of) the terms involving the Goldstone boson should be resummed to all orders, roughly speaking replacing its mass-squared parameter (which appears in the loop functions) with the equivalent parameter derived from the total effective potential (i.e. zero, since it is a Goldstone boson). In section \ref{SEC:tadpoles} of this work we show how this can be extended to general renormalisable theories. Note that this approach is related to the (symmetry-improved) two-particle-irreducible potential approach pioneered in \cite{Pilaftsis:2013xna,Pilaftsis:2015cka,Pilaftsis:2015bbs}, where essentially all particle propagators are resummed -- which is somewhat more difficult to automate. 
 
In \cite{Martin:2014bca,Elias-Miro:2014pca} it was noted that the Goldstone resummation would not regulate divergences in the second derivatives of the effective potential, and so to have a divergence-free calculation of the neutral scalar (i.e. Higgs) masses it would be necessary to include the external momentum in the self energies rather than using an effective potential approximation. This is particularly important because the zero momentum approximation is widely used to calculate the Higgs mass \cite{Hempfling:1993qq, Heinemeyer:1998jw, Heinemeyer:1998kz,
  Zhang:1998bm, Heinemeyer:1998np, Espinosa:1999zm, Espinosa:2000df,
  Degrassi:2001yf, Brignole:2001jy, Brignole:2002bz, Martin:2002iu,
  Martin:2002wn, Dedes:2003km, Heinemeyer:2004xw,Martin:2007pg, Harlander:2008ju,Degrassi:2009yq,
  Kant:2010tf,Goodsell:2014bna,Dreiner:2014lqa,Muhlleitner:2014vsa,Goodsell:2015ira,Nickel:2015dna,Staub:2015aea,Diessner:2015yna,Goodsell:2015yca,Diessner:2015iln,Goodsell:2016udb,Braathen:2016mmb} -- indeed there are few publicly available implementations of diagrammatic calculations of the Higgs mass beyond one loop in theories beyond the Standard Model which do not use it (some momentum-dependent diagrammatic calculations are available for the MSSM \cite{Martin:2004kr,  Borowka:2014wla, Degrassi:2014pfa}). While the Goldsone Boson Catastrophe can be avoided in the MSSM in the gaugeless limit (where the Goldstone boson does not couple to the Higgs, and so generates no divergences) it is of pressing concern for more general theories, since the two-loop computation has recently become publicly available through \SARAH \cite{Staub:2008uz,Staub:2009bi,Staub:2010jh,Staub:2012pb,Staub:2013tta,Staub:2015kfa,Goodsell:2014bna,Goodsell:2015ira}; the Goldstone Boson Catastrophe as it affects that implementation was discussed in \cite{Goodsell:2014pla,Goodsell:2016udb}, and recently manifested itself in \cite{Goodsell:2015ura,Benakli:2016ybe,Athron:2016fuq}. Indeed, while the numerical impact of the problem in the Standard Model seems to be small (at least away from the divergent points, simply neglecting the imaginary part of the potential seems to give results close to those of the full solutions), in more complicated theories it can cause divergent contributions to the masses for many regions of the parameter space; in \cite{Goodsell:2015ura,Benakli:2016ybe} it was necessary to restrict to only the two-loop corrections proportional to the strong gauge coupling for those regions in performing parameter scans.

In section \ref{SEC:masses}, we shall show that the inclusion of external momentum in the scalar self-energies does \emph{not} by itself avoid all divergences. In fact, it is necessary to resum the Goldstone boson contributions in the mass diagrams too -- to cancel the divergences in a class of diagrams which do not depend on momentum. We will also show that the resummation can be implemented most easily to two loop order by using an ``on-shell'' scheme for the Goldstone bosons. With these modifications, to cure the remaining divergences the diagrammatic implementation in \cite{Goodsell:2015ira} could in principle be extended to include the external momentum by changing the loop functions to those implemented in {\tt TSIL} \cite{Martin:2003qz,Martin:2005qm}. However, analytic expressions for general loop functions with momenta are not known: they are in general obtained by solving differential equations, which is numerically expensive. Therefore, in appendix \ref{SEC:momentum} we give a complete set of analytic expressions for expansions of the necessary functions including all divergent and constant terms in an expansion of the four-momentum-squared $s$ around zero (neglecting those of $\mathcal{O}(s)$). This allows fast evaluation of a generalised effective potential approximation for the neutral scalar masses -- although for this part we shall be restricted to the gaugeless limit (setting the couplings of all broken gauge groups to zero) since the mass diagrams are known only up to second order in the gauge couplings.  

Once the Goldstone Boson Catastrophe has been solved, using similar techniques it was shown in \cite{Martin:2014bca,Elias-Miro:2014pca,Kumar:2016ltb} that it is also possible to improve the solution of the tadpole equations for the other mass-squared parameters (not just the one corresponding to the tree-level Goldstone boson mass). In general, the same mass-squared parameters $m^2$ appear both as solutions of the tadpole equations, and in the loop functions, in the schematic form
$$m^2 = m_0^2 -\frac{1}{v} \frac{\partial \Delta V (m^2)}{\partial v}$$
where $m_0^2$ is the tree-level solution of the tadpole equation, $v$ is some expectation value and $ \Delta V$ are the loop corrections to the effective potential. 
Although resummation is not required for them (except perhaps for the Higgs boson, where the quantum corrections are so large that they force its tree-level mass to become negative -- we shall not discuss such a case here), these other mass-squared parameters can be expanded perturbatively in the loop functions so that the equations can be solved directly rather than iteratively. In other words, we find only the tree-level values of the parameters on the right-hand side of the equation, and the loop-corrected solution on the left (as opposed to the loop-corrected value on both sides):
$$m^2 = m^2_0 -\frac{1}{v} \frac{\partial \Delta V (m^2_0)}{\partial v} - \delta\bigg(\frac{1}{v}\frac{\partial \Delta V (m^2_0)}{\partial v} \bigg)\ ;$$
we shall refer to these throughout as ``self-consistent equations''. In section \ref{SEC:reexpansion}, we will show how to carry out this procedure in general, showing that the formulae can be given in simpler form than in, e.g., \cite{Kumar:2016ltb} for the MSSM case. We shall also go further and show how this shifts the mass diagrams.

Finally, we have endeavoured to keep the paper as self-contained as possible, and for that purpose we provide in appendix \ref{SEC:loopfn} a set of all of the loop functions used throughout.

\section{The Goldstone Boson Catastrophe and resummation}
\label{SEC:GBC}

\subsection{Abelian Goldstone model}
\label{SUBSEC:example}

Let us begin by recalling the problem of the Goldstone Boson Catastrophe. For simplicity we shall take the simplest abelian Goldstone model  defined by a complex scalar field $\Phi$ (and no gauge group) with potential 
\begin{align}
V =& \mu^2 |\Phi|^2 + \lambda |\Phi|^4
\end{align}
and expand around an expectation value $v$ as $\Phi = \frac{1}{\sqrt{2}} (v + h + i G)$ to obtain
\begin{align}
\vtree =& \frac{v^2}{4} (\lambda v^2 + 2 \mu^2) + hv(v^2\lambda + \mu^2)  + \frac{1}{2} (3 v^2 \lambda + \mu^2) h^2 + \frac{1}{2} (\mu^2 + \lambda v^2) G^2 \nn\\
&+ v \lambda ( h^3 + h G^2) + \frac{\lambda}{4} (h^4 + 2 G^2 h^2 + G^4). 
\end{align}
Defining $m_G^2 \equiv \mu^2 + \lambda v^2, M_h^2 \equiv \mu^2 + 3 v^2 \lambda, $
we can then compute the effective potential up to two loops:
\begin{align}
V_\text{eff}(v) \equiv & V^{(0)}\big|_{h,G=0} + \frac{1}{16\pi^2} V^{(1)} + \frac{1}{(16\pi^2)^2}V^{(2)} + ... \nn\\
= & \vtree\big|_{h,G=0} + \frac{1}{16\pi^2} ( f(m_G^2) + f(M_h^2)) + \frac{\lambda}{(16\pi^2)^2} \bigg[ \frac{3}{4} A(m_G^2)^2 + \frac{1}{2} A(m_G^2) A(M_h^2) + \frac{3}{4} A(M_h^2)^2 \bigg] \nn\\
&  - \frac{\lambda^2 v^2}{(16\pi^2)^2} \bigg[ 3 I(M_h^2, M_h^2,M_h^2) + I(M_h^2, m_G^2,m_G^2)\bigg] + ...
\label{abelian_goldstone_veff}\end{align}
where 
the one-loop functions $f(x)$, $A(x)$
and the two-loop function $I(x,y,z)$ are defined in the appendix, equations~(\ref{expr_f}),~(\ref{expr_A}), and~(\ref{expr_I}). 
The potential is regular as $m_G \rightarrow 0$ but does contain terms of order $m_G^2 \blog m_G^2$ (where $\blog x$ is also defined in the appendix -- eq.~(\ref{llog_def})) so that when we derive the tadpole equation and expand the derivative of $I(M_h^2, m_G^2, m_G^2)$ around $m_G^2 = 0$ we find
\begin{align}
0 =& \frac{\partial V_\text{eff}}{\partial v} = m_G^2 v + \frac{2\lambda v}{16\pi^2} \bigg[ \frac{1}{2} A(m_G^2) + \frac{3}{2} A(M_h^2) \bigg] \nn\\
& + \frac{2\lambda^2 v}{(16\pi^2)^2} \llog m_G^2 \bigg[ \frac{3}{2} A(m_G^2)   + \frac{1}{2} A(M_h^2) + \frac{2\lambda v^2}{M_h^2} A(M_h^2) \bigg]  \nn\\
&+ \mathrm{other\ non-singular\ terms.}
\end{align}
 The $\llog m_G^2$ terms on the second line are the manifestation of the Goldstone Boson Catastrophe: we cannot insert the tree-level solution $m_G^2 =0$ into them, and will have a complex potential if we find $m_G^2 < 0$. The solution proposed in \cite{Martin:2014bca,Elias-Miro:2014pca} is to resum the Goldstone boson propagators -- in the one-loop effective potential we make the substitution
\begin{align}
V_\text{eff}^{(1)} \supset& -\frac{i}{2} C \int d^d k \log (-k^2 + m_G^2) \rightarrow  -\frac{i}{2} C\int d^d k \log (-k^2 + m_G^2 + \Pi_{GG} (k^2)) \nn\\
\rightarrow&  -\frac{i}{2} C\int d^d k \log (-k^2 + m_G^2 + \Pi_{GG} (0 )) + ...
\label{EQ:Veff1}\end{align}
where $C$ is a constant defined in equation (\ref{EQ:definitionC}), and $\Pi_{GG}(k^2) $ is the Goldstone boson self energy, given here 
at one loop by
\begin{align}
\Pi_{GG} (k^2) =& \frac{1 }{16\pi^2} \bigg[3\lambda A(m_G^2) + \lambda A(M_h^2) - 4 \lambda^2 v^2 B(k^2,m_G^2, M_h^2) \bigg].
\end{align}
With zero external momentum, this becomes
\begin{align}
 \Pi_{GG} (0)  =&\frac{1 }{16\pi^2} \bigg[\lambda A(m_G^2) + 3\lambda A(M_h^2)\bigg].
\end{align}
The term involving only the Goldstone mass-squared will not have a well-defined derivative, and this also leads to divergences  when we resum the effective potential at three loops and above. The prescription of \cite{Elias-Miro:2014pca} is to drop it in favour of $\Pi_g = \frac{1}{16\pi^2}\Pi_g^{(1)} + ... $ where
\begin{align}
\Pi_{g}^{(1)} (0) =&   3\lambda A(M_h^2)=\lambda A(M_h^2)-4\lambda^2 v^2 B(0,0, M_h^2).
\end{align}
Note that this does not correspond to dropping one particular class of diagrams (at one loop it is a combination of the one- and two-propagator diagrams) but instead must be defined in terms of dropping contributions from ``soft'' Goldstone bosons.  
Defining 
\begin{align}
\Delta \equiv \Pi_g (0) \equiv \frac{1 }{16\pi^2}\Delta_1 + \frac{1 }{(16\pi^2)^2}\Delta_2 + \cdots ,
\label{EQ:DefDelta}\end{align}
we then should use instead the resummed potential
\begin{align}
\veffr \equiv& V_\text{eff} + \frac{1}{16\pi^2} \bigg[ f( m_G^2 + \Delta) - \sum_{n=0}^{l-1} \frac{\Delta^n}{n !} \left(\frac{\partial }{\partial m_G^2}\right)^n f(m_G^2) \bigg] 
\end{align}
where $l$ is the loop order to which $V_\text{eff}$ has been calculated; the terms in square brackets simply ensure that the potentials are identical up to $l$ loops and only differ at higher orders. Performing this procedure for the potential above we find
\begin{align}
\veffr =& \vtree + \frac{1}{16\pi^2} ( f(m_G^2 + \Delta) + f(M_h^2))  + \frac{\lambda}{(16\pi^2)^2} \bigg[ \frac{3}{4} A(m_G^2)^2  + \frac{3}{4} A(M_h^2)^2 \bigg] \nn\\
& - \frac{\lambda^2 v^2}{(16\pi^2)^2} \bigg[ 3 I(M_h^2, M_h^2,M_h^2) + I(M_h^2, m_G^2,m_G^2)+\frac{1}{\lambda v^2} A(M_h^2)A(m_G^2)\bigg] .
\end{align}
With the above procedure, we have resummed the leading divergences at two loops, i.e. the terms of order $m_G^2 \llog m_G^2 $ for small $m_G^2$ (we expect $m_G^2$ it to be of order a one-loop quantity at the minimum). If we are interested in the first derivative of the potential then this is sufficient; to find the minimum to two-loop order we can expand the potential to order $m_G^2$ with the help of eq.~(\ref{exp_fsss_2g}):
\begin{align}
\label{first_veffr}
\veffr =& \vtree + \frac{1}{16\pi^2} ( f(m_G^2 + \Delta) + f(M_h^2)) \\
& + \frac{\lambda}{(16\pi^2)^2} \bigg[ \frac{3}{4} A(M_h^2)^2 \bigg] - \frac{\lambda^2 v^2}{(16\pi^2)^2} \bigg[ 3 I(M_h^2, M_h^2,M_h^2) + I(M_h^2, 0,0)-2R_{SS}(0, M_h^2)m_G^2\bigg] + \mathcal{O}(m_G^4), \nn
\end{align}
making the regularity apparent, although note that the higher-order terms still contain a $m_G^4 \llog m_G^2$ term. 
The tadpole equation, neglecting terms of three-loop order, is then
\begin{align}
\label{EQ:dVeffEsp}
 0 =& v \left(m_G^2 +\frac{1}{16\pi^2} \Delta_1\right) + \frac{\lambda v+ \Delta^\prime/2}{16\pi^2} A(m_G^2 + \Delta) \\
 &+\frac{1}{(16\pi^2)^2}\bigg\{\lambda\bigg[9\lambda vA(M_h^2)\llog M_h^2\bigg]-2\lambda^2 v \bigg[ 3 I(M_h^2, M_h^2,M_h^2) + I(M_h^2,0,0)\bigg] \nn\\
 &\quad\quad\quad\quad\quad+\lambda^2v^2\bigg[6\lambda v\left(9U_0(M_h^2,M_h^2, M_h^2,M_h^2)+U_0(M_h^2,M_h^2,0,0)\right)+4\lambda vR_{SS}(0,M_h^2)\bigg]\bigg\}. \nn
\end{align}
Noting that the solution to the one-loop equation is $m_G^2 + \frac{1}{16\pi^2}\Delta_1 = 0 $, we see that we can neglect the $A(m_G^2 + \Delta)$ term as it gives a correction of order three loops. We ought then to find that we can identify the term in curly brackets with $\Delta_2$: for a Goldstone boson we should find $m_G^2 + \Pi_{GG} (0) =0,$
so we expect that we should be able in general to identify $\frac{1}{v} \frac{\partial  V^{(\ell)}}{\partial v} = \Pi_{GG}^{(\ell)} (0)$, and therefore for our modified potential we should expect
\begin{align}
\frac{1}{v} \frac{\partial  \hat{V}^{(\ell)}}{\partial v} =& \Pi_g^{(\ell)} (0).
\end{align}
This leads to the prescription in \cite{Martin:2014bca,Kumar:2016ltb}, which is somewhat simpler: we expand the potential $\veff$ as a series in $m_G^2$:
\begin{equation}
 \vtwo=\vtwo|_{m_G^2=0}+\frac{1}{2}\Delta_1A(m_G^2)+\frac{1}{2}\Omega m_G^2+\mathcal{O}(m_G^4).
\end{equation}
We can then use this as the \emph{definition} of $\Delta_1$ instead of equation (\ref{EQ:DefDelta}). We then resum the effective potential  as
\begin{align}
 \veffr=&\vtree+\frac{1}{16\pi^2}\bigg[\vone|_{m_G^2=0}+f\left(m_G^2+\Delta\right)\bigg]+\frac{1}{(16\pi^2)^2}\bigg[\vtwo|_{m_G^2=0}+\frac{1}{2}\Omega m_G^2\bigg].
\end{align}
By doing this, we immediately find the expression in (\ref{first_veffr}), with $\Omega \equiv 4\lambda^2 v^2 R_{SS}(0, M_h^2) $.
When we take the derivative and expand up to two-loop order then the minimum is at $m_G^2 + \Delta =0$ with
\begin{align}
\Delta_2 =&
\bigg[ \frac{1}{v}\frac{\partial}{\partial v} \left(\vtwo|_{m_G^2=0}\right) + \lambda \Omega \bigg].
\end{align}
We shall follow this second procedure to find the minimum condition in general renormalisable field theories at two loop order. 

We shall also consider a hybrid approach, which is to adopt an \emph{on-shell} condition for the Goldstone boson: we define 
\begin{align}
(m_G^2)^{\rm dim.\ reg} \equiv& (m_G^2)^{\text{OS}} - \Pi_{GG} ((m_G^2)^{\text{OS}}) = - \Pi_{GG} (0).
\end{align} 
This is particularly effective at two loops, where we only need $\Pi_{GG}^{(1)}$; furthermore, since $ (m_G^2)^{\text{OS}} =0,$ at this loop order there is no difference between $\Pi_{GG}$ and $\Pi_g$. Making the above substitution in the potential we find exactly the same result as our resummed version in equation (\ref{first_veffr}). However, we also have the advantage that we can make this substitution directly \emph{in the tadpole equation}:
\begin{align}
0 =& v(m_G^2)^{\rm dim.\ reg} + \frac{\lambda v}{16\pi^2} A((m_G^2)^{\rm dim.\ reg}) + \frac{3\lambda v}{16\pi^2} A(M_h^2)  + \frac{1}{(16\pi^2)^2}\frac{\partial V^{(2)}}{\partial v} \\
=&  v(m_G^2)^{\rm dim.\ reg} + \frac{3\lambda v}{16\pi^2} A(M_h^2)  + \frac{1}{(16\pi^2)^2}\lim_{(m_G^2)^{\text{OS}}\rightarrow 0} \bigg[\frac{\partial V^{(2)}}{\partial v}((m_G^2)^{\text{OS}} ) - 3\lambda^2 v A(M_h^2) \blog((m_G^2)^{\text{OS}} )\bigg],\nn
\end{align}
which gives \emph{exactly} the expression that we found above in (\ref{EQ:dVeffEsp}). We shall find in the following that this simple approach is also exactly what we need for the \emph{mass} diagrams. However, we must first introduce some notation and formalism to handle the general case when (potentially several) Goldstone bosons and neutral scalars can mix.

\subsection{Notation for general field theories}
\label{SUBSEC:notations}

In the previous subsection we considered the simplest possible model where there were only two real scalars which cannot mix. Once we consider more general theories, there can be more Goldstone bosons and, even when they have been identified at tree level, they can in general mix with other scalars (only pseudoscalars in the case of CP conservation) once loop corrections are included. This problem does not arise in the Standard Model as treated in references \cite{Martin:2014bca,Elias-Miro:2014pca}, because all of the pseudoscalars are would-be Goldstone bosons and the neutral and charged Goldstones cannot mix, so can be treated as two separate sectors. In the MSSM, there are additional scalars and pseudoscalars, but in the CP-conserving case considered in \cite{Kumar:2016ltb} the mixing is at most among \emph{pairs} of fields, and could be written in each case in terms of mixing angles and $2\times 2$ matrices. Furthermore, the same applies for all of the scalars treated so far in those references: at most pairs of fields could mix. These complications are particularly important because in the previous cases the simplest way to derive the tadpole equations was to write down the potential and take the derivatives; once we consider more complicated cases this is no longer true and we will want to be able to directly write down expressions for the derivatives as in \cite{Goodsell:2015ira}.

Starting with the scalar sector, since we will need to take the derivatives of the potential with respect to scalar fields, we shall follow the procedure in \cite{Goodsell:2015ira}. We introduce first the unrotated scalar potential in terms of real scalar fields  $\varphi^0$ and their fluctuations around expectation values $v_i$ such that $\varphi^0_i \equiv v_i + \phi^0_i$:
\begin{equation}
\label{scalar_pot}
 \vtree(\{\varphi_i^0\})=\vtree(v_i)+\frac{1}{2}\hat{m}_{0,ij}^2\phi_i^0\phi_j^0+\frac{1}{6}\hat{\lambda}_0^{ijk}\phi_i^0\phi_j^0\phi_k^0+\frac{1}{24}\hat{\lambda}_0^{ijkl}\phi_i^0\phi_j^0\phi_k^0\phi_l^0.
\end{equation}
Here $\hat{m}_{0,ij}^2 $ satisfies the tree-level tadpole equations. 
From this we can define the field-dependent masses and couplings, 
\begin{align}
\hat{m}_{ij}^2(\phi^0)&\equiv\frac{\partial^2\vtree}{\partial\phi_i^0\partial\phi_j^0}=\hat{m}_{0,ij}^2+\hat{\lambda}^{ijk}_0\phi_k^0+\frac{1}{2}\hat{\lambda}_0^{ijkl}\phi_k^0\phi_l^0,\\
\hat{ \lambda}^{ijk}(\phi^0)&\equiv\frac{\partial^3\vtree}{\partial\phi_i^0\partial\phi_j^0\partial\phi_k^0}=\hat{\lambda}^{ijk}_0+\hat{\lambda}^{ijkl}_0\phi_l^0,\\
 \hat{\lambda}^{ijkl}(\phi^0)&\equiv\frac{\partial^4\vtree}{\partial\phi_i^0\partial\phi_j^0\partial\phi_k^0\partial\phi_l^0}=\hat{\lambda}_0^{ijkl}.
\end{align} 
We then introduce a new basis $\{\tilde{\phi}_i\}$ and an orthogonal matrix $\Rt$ to diagonalise the tree-level mass matrix as 
\begin{equation} 
 \phi_i^0=\tilde{R}^{}_{ij}\tilde{\phi}_j,
\end{equation}
and obtain the new masses and couplings
\begin{align}
\label{fielddep_couplings}
 \tilde{m}_i^2\delta_{ij}&=\hat{m}_{kl}^2\tilde{R^{}}_{ki}\tilde{R^{}}_{lj}\\
 \tilde{\lambda}^{ijk}&=\hat{\lambda}^{lmn}\tilde{R^{}}_{li}\tilde{R^{}}_{mj}\tilde{R^{}}_{nk}\\
 \tilde{\lambda}^{ijkl}&=\hat{\lambda}^{mnpq}\tilde{R^{}}_{mi}\tilde{R^{}}_{nj}\tilde{R^{}}_{pk}\tilde{R^{}}_{ql}.
\end{align}

Next we need to define what happens when we introduce the loop corrections to the effective potential $\Delta V$ and modify the tadpole equations. We shall take the expectation values $v_i$ to be fixed (i.e. they are the true values at the minimum of the full quantum-corrected potential) and instead correct the mass-squared parameters in the lagrangian, passing from $ \hat{m}_{0,ij}^2 $ (which satisfy the tree-level tadpole equations) to new quantities $m_{ij}^2$. Using the minimisation conditions,  the relationship between them is 
\begin{align}
m_{ij}^2 v_j = \hat{m}_{0,ij}^2 v_j - \frac{\partial \Delta V}{\partial \phi^0_i}\bigg|_{\phi^0_i = 0} .
\label{EQ:masssqeq}\end{align}
Diagonalising these requires the introduction of a new basis via $\phi^0_i = R_{ij} \phi_j$, having masses $m_i$ and couplings $\lambda^{ijk}, \lambda^{ijkl}$.

For the couplings involving fermions and scalars, we shall use the notation for a general renormalisable field theory used in \cite{Martin:2001vx,Martin:2003it}; we repeat here the scalar, scalar-fermion and scalar-gauge-boson interactions:
\begin{align}
\lagr_S =& -\frac{1}{6} \lambda^{ijk} \phi_i \phi_j \phi_k - \frac{1}{24} \lambda^{ijkl} \phi_i \phi_j \phi_k \phi_l,\nn\\
\lagr_{SF} =& - \frac{1}{2} y^{IJk} \psi_I \psi_J \phi_k + c.c., \nn\\
\lagr_{SV} =& - \frac{1}{2} g^{abi} A_\mu^a A^{\mu b} \phi_i - \frac{1}{4} g^{abij} A_\mu^a A^{\mu b} \phi_i \phi_j - g^{aij} A_\mu^a \phi_i \partial^\mu \phi_j.
\label{EQ:couplingdefinitions}\end{align}
The fermions here are in Weyl notation and are supposed to be defined in a basis where the masses of all fields are diagonal.

\subsection{Goldstone bosons in general field theories}
\label{SUBSEC:goldstone}

To deal with Goldstone boson mixing in general theories, we will need some notation and simple results. We start from a theory with a global symmetry such that the scalars transform under a set of infinitesimal shifts as $\phi_i \rightarrow \phi_i + \epsilon^G \alpha^G_i$. Then the standard result is to expand $V(\phi_i + \epsilon^G \alpha^G_i) = V(\phi_i)$ and differentiate the relation once:
\begin{align}
\epsilon^G \alpha^G_i \frac{\partial V}{\partial \phi_i^0} = 0, \qquad \frac{\partial (\epsilon^G \alpha^G_i)}{\partial \phi_j^0} \frac{\partial V}{\partial \phi_i^0} + \epsilon^G \alpha^G_i \frac{\partial^2 V}{\partial \phi_i^0 \partial \phi_j^0} =0.
\end{align}
When we sit at the minimum of the potential $\frac{\partial V}{\partial \phi_i^0} =0 $ but for a spontaneously broken symmetry $\alpha_i^G$ is not zero for all $i$, and thus we have a null eigenvector of the scalar mass matrix -- i.e. the Goldstone boson. For more than one symmetry broken then there will be multiple null eigenvectors and these should be formed into an orthonormal set. Let us write the symmetry shifts as linear coefficients $\alpha_i^G = a_{ij}^G \phi_j^0$ after this has been performed so that $\sum_i \alpha_i^G \alpha_i^{G^\prime} = \delta^{G G^\prime} $ and then
\begin{equation}
\label{goldstone_def}
 G_G=\phi_G=R^{}_{jG}\phi_j^0\text{, where }R^{}_{jG}=\alpha^G_j .
\end{equation}
We use the index ``$G$'' now to refer to the Goldstone boson(s) in the diagonal basis. The first identity that we need arises from taking a further derivative of the above equations to give
\begin{align}
\label{EQ:no_3G_coupling}
\epsilon^G \alpha^G_i \frac{\partial^3 V}{\partial \phi_i^0 \partial \phi_j^0 \partial\phi_k^0} +  \frac{\partial^2 (\epsilon^G \alpha^G_i)}{\partial \phi_j^0 \partial \phi_k^0} \frac{\partial V}{\partial \phi_i^0} +  \frac{\partial (\epsilon^G \alpha^G_i)}{\partial \phi_j^0} \frac{\partial^2 V}{\partial \phi_i^0 \partial \phi_k^0} +  \frac{\partial (\epsilon^G \alpha^G_i)}{\partial \phi_k^0} \frac{\partial^2 V}{\partial \phi_i^0 \partial \phi_j^0} = 0 \nn\\
\rightarrow \alpha^G_i \alpha_j^{G^\prime} \alpha_k^{G^{\prime\prime}}   \frac{\partial^3 V}{\partial \phi_i^0 \partial \phi_j^0 \phi_k^0}\bigg| =& 0,
\end{align}
i.e. there are no three-Goldstone couplings.

If we were able to work at the true minimum of the potential and with self-consistent values of all the parameters then this would be sufficient. However, we must use the minimum conditions to determine the parameters -- a subset of the mass-squared parameters, in our case -- and this means that the above equations will be violated by loop corrections. In particular, the  mass-squared parameter -- in the diagonal basis -- for the would-be Goldstone boson is no longer zero. To see this, let us define the loop tadpoles 
\begin{align}
\delta_i \equiv& \frac{1}{v_i} \frac{\partial \Delta V}{\partial \phi^0_i} \bigg|_{\phi_i^0 = 0}
\end{align}
so that we can solve (\ref{EQ:masssqeq}) with the commonly-made choice of
\begin{align}
m_{ij}^2 =& - \delta_i \delta_{ij}+ \hat{m}_{0,ij}^2 .
\label{EQ:tadpolesolution}\end{align}
Note that this is the value at the minimum of the potential -- so $\delta_i$ is \emph{not} regarded as a function of $\{\phi^0_i\}$ when we take derivatives below.
Now
\begin{align}
m_G^2 =& (R^T m^2 R)_{GG} = - \sum_i \Rt_{iG}^2 \delta_i + \mathcal{O} ({\rm 2\ loops}),
\end{align}
i.e. we can use the tree-level rotation matrices to obtain the Goldstone mass from the loop tadpoles up to corrections of two-loop order, which is all we shall require in the following. This generalises, for example, equations (2.26) and (2.27) of \cite{Kumar:2016ltb}.

Following equation~(\ref{EQ:no_3G_coupling}) above, we then see that 
\begin{align}
\tilde{\lambda}^{GG'G''}=0,\qquad \lambda^{GG'G''}  = \mathcal{O}({\rm 1\ loop})
\label{EQ:goldstoneselfcoupling}\end{align}
in general. This is a crucial result in the following, even if in theories that preserve CP both couplings are zero to all orders. For theories breaking CP that could generate such a term at one or two loops, when we expand the potential as a series in $m_G^2$ as in section \ref{SUBSEC:example} (justified by it being a one-loop quantity) we shall \emph{also implicitly expand the Goldstone self-coupling $\lambda^{GG'G''}$} for the same reason; implicitly because we shall not need the higher-order terms and this just corresponds to setting $\lambda^{GG'G''} =0 $ everywhere. Note that this is automatic once we also employ re-expansion of the tadpoles and masses in terms of tree-level parameters to obtain consistent tadpole equations in section \ref{SEC:reexpansion}. 

In practice when we are considering the broken gauge groups to be $SU(2) \times U(1)_Y$ the unbroken $U(1)_{\rm QED}$  allows the Goldstones to be separated into one neutral and one (complex) charged Goldstone that cannot mix. Hence in the following to simplify the notation we will restrict to a single neutral Goldstone boson and drop the lower index $G$, but the treatment of the charged Goldstone is identical. In this case we can also write $\epsilon^G \alpha_i^G \rightarrow a_{ij} \phi_j^0$ and thus $R_{jG} = \frac{a_{ij} v_j}{\sqrt{a_{ij} a_{ik} v_jv_k}}$ (where we now allow the normalisation of $a_{ij} v_j$ to be arbitrary) for the linearly realised symmetries considered here.

\subsection{Small $\gbm$ expansion of the effective potential for general theories}
\label{SUBSEC:expansion}

To close this section we can now apply the notation and machinery from the previous subsections to resum the general effective potential at two loops, generalising the procedure of \cite{Martin:2014bca,Elias-Miro:2014pca}. 

The total potential up to two loops expands as 
\begin{align}
 \veff=\vtree+\frac{1}{16\pi^2}\vone+\frac{1}{(16\pi^2)^2}\vtwo.
\end{align}
For use in the elimination of the infrared divergences in the derivatives of the effective potential, we expand $\veff$ for small $\gbm$. More precisely, we 
want to write the two-loop part of $\veff$ as 
\begin{equation}
\label{exp_2l} 
 \vtwo=\vtwo|_{m_G=0}+\frac{1}{2}A(m_G^2)\Delta_1+\frac{1}{2}m_G^2\Omega + \mathcal{O}(m_G^4),
\end{equation}
where the quantities $\Delta_1$ and $\Omega$ are to be determined.

The two-loop potential splits into contributions \cite{Martin:2001vx}: 
\begin{align}
\vtwo =& \vtwo_{SSS}+\vtwo_{SS} + \vtwo_{FFS}+\vtwo_{\ov{FF}S} + \vtwo_{SSV} +  \vtwo_{SV} + \vtwo_{VVS} + \bigg(  \vtwo_{FFV}+\vtwo_{\ov{FF}V} +\vtwo_{\rm gauge} \bigg)
\end{align}
where the subscripts denote the propagators in the loops as scalar, fermion or vector (gauge sector). The terms in the brackets will not be resummed (since they contain no scalars) and so can be taken to be unchanged from the expressions in \cite{Martin:2001vx}. The loop functions appearing in the other terms are recalled in the \msbar and \drbar schemes and Landau gauge in appendix \ref{SUBSUBSEC:loopfndef_2l}. 

First, the scalar contributions to the effective potential at two-loop order $\vtwos\equiv  \vtwo_{SSS} + \vtwo_{SS}$  read 
\begin{align}
 \vtwo_{SSS}&\equiv\frac{1}{12}(\lambda^{ijk})^2f_{SSS}(m_i^2,m_j^2,m_k^2),\\
 \vtwo_{SS}&\equiv\frac{1}{8}\lambda^{iijj}f_{SS}(m_i^2,m_j^2),
\end{align}
and these functions can be expanded using formulae (3.7), (3.8) of \cite{Kumar:2016ltb}. Separating terms with one or more Goldstone bosons from the terms 
without any, and using the fact that $\lambda^{GGG}$ vanishes at leading order -- see the discussion around equation (\ref{EQ:goldstoneselfcoupling}) -- we find the expansion of $\vtwos$:
\begin{align}
\label{EQ:vtwo_scalar}
\vtwos=&\,\vtwos|_\text{no GB}+\sum_{j,k\neq G}\frac{1}{4}(\lambda^{Gjk})^2f_{SSS}(0,m_j^2,m_k^2)+\sum_{k\neq G}\frac{1}{4}(\lambda^{GGk})^2f_{SSS}(0,0,m_k^2)\nn\\
&+A(m_G^2)\bigg(\sum_{j,k\neq G}\frac{1}{4}(\lambda^{Gjk})^2P_{SS}(m_j^2,m_k^2)+\sum_{j\neq G}\frac{1}{4}\lambda^{GGjj}A(m_j^2)
+\sum_{k\neq G}\frac{1}{2}(\lambda^{GGk})^2P_{SS}(0,m_k^2)\bigg)\nn\\
&+m_G^2\bigg(\sum_{j,k\neq G}\frac{1}{4}(\lambda^{Gjk})^2R_{SS}(m_j^2,m_k^2)+\sum_{k\neq G}\frac{1}{2}(\lambda^{GGk})^2R_{SS}(0,m_k^2)
\bigg)+\mathcal{O}(m_G^4),
\end{align}
from which we can identify the scalar part of $\Delta_1$ and $\Omega$
\begin{align}
\label{delta1s}
 &(\Delta_1)_S=\,\sum_{j,k\neq G}\frac{1}{2}(\lambda^{Gjk})^2P_{SS}(m_j^2,m_k^2)+\sum_{j\neq G}\frac{1}{2}\lambda^{GGjj}A(m_j^2)+\sum_{k\neq G}(\lambda^{GGk})^2P_{SS}(0,m_k^2),\nn\\
 &\Omega_S=\,\sum_{j,k\neq G}\frac{1}{2}(\lambda^{Gjk})^2R_{SS}(m_j^2,m_k^2)+\sum_{k\neq G}(\lambda^{GGk})^2R_{SS}(0,m_k^2).
\end{align}

Next, the terms in $\vtwo$ involving fermions and scalars are 
\begin{align}
 \vtwo_{FFS}&\equiv\frac{1}{2}y^{IJk}y_{IJk}f_{FFS}(m_I^2,m_J^2,m_k^2),\\
 \vtwo_{\ov{FF}S}&\equiv\frac{1}{2}\text{Re}\bigg[y^{IJk}y^{I'J'k}M^*_{II'}M^*_{JJ'}\bigg]f_{\ov{FF}S}(m_I^2,m_J^2,m_k^2).
\end{align}
Here, there are only two cases to consider, either $k\neq G$ or $k=G$, and for the latter case, we can use eqs. (3.9) and (3.10) 
from \cite{Kumar:2016ltb} to expand the loop functions for small $\gbm$. We then obtain for $(\Delta_1)_{FS}$ and $\Omega_{FS}$
\begin{align}
 &(\Delta_1)_{FS}=y^{IJG}y_{IJG}P_{FF}(m_I^2,m_J^2)+\text{Re}\bigg[y^{IJG}y^{I'J'G}M^*_{II'}M^*_{JJ'}\bigg]P_{\ov{F}\ov{F}}(m_I^2,m_J^2),\\
 &\Omega_{FS}=y^{IJG}y_{IJG}R_{FF}(m_I^2,m_J^2)+\text{Re}\bigg[y^{IJG}y^{I'J'G}M^*_{II'}M^*_{JJ'}\bigg]R_{\ov{F}\ov{F}}(m_I^2,m_J^2).
\end{align}

Finally, the terms with scalars and gauge bosons read
\begin{align}
\vtwo_{SSV}=&\frac{1}{4}(g^{aij})^2f_{SSV}(m_i^2,m_j^2,m_a^2),\\
\vtwo_{VS}=&\frac{1}{4}g^{aaii}f_{VS}(m_a^2,m_i^2),\\
\vtwo_{VVS}=&\frac{1}{4}(g^{abi})^2f_{VVS}(m_a^2,m_b^2,m_i^2).
\end{align}
As previously, we can expand these terms and separate the contributions of the Goldstone boson, and we find
\begin{align}
 &(\Delta_1)_{VS}=\frac{3}{2}g^{aaGG}A(m_a^2)+\frac{1}{2}(g^{abG})^2P_{VV}(m_a^2,m_b^2),\\
 &\Omega_{VS}=(g^{aGj})^2R_{SV}(m_j^2,m_a^2)+(g^{aGG})^2R_{SV}(0,m_a^2)+\frac{1}{2}(g^{abG})^2R_{VV}(m_a^2,m_b^2).
\end{align}

The expansion~(\ref{exp_2l}) of $\vtwo$ enables us to rewrite the two-loop effective potential after resummation of the leading Goldstone boson contributions as 
\begin{align}
\label{resummed_veff}
 \veffr=&\vtree+\frac{1}{16\pi^2}\bigg(\vone|_{m_G^2=0}+f(\gbm+\Delta_G)\bigg)+\frac{1}{(16\pi^2)^2}\bigg(\vtwo|_{m_G^2=0}+\frac{1}{2}\Omega\, m_G^2\bigg), \\
\Omega =& \Omega_S + \Omega_{FS} + \Omega_{VS}, \quad \Delta_G \equiv \sum_{i} R_{iG}^2 \frac{1}{v_i} \frac{\partial \hat{V}_{\rm eff}}{\partial \phi_i^0} = \frac{1}{16\pi^2} \bigg[(\Delta_1)_S+(\Delta_1)_{FS}+(\Delta_1)_{VS}\bigg] + \Otl .\nn
\end{align}
The minimum of this potential will be found at $m_G^2 + \Delta_G = 0$ (along with the minimisation conditions for the additional scalars) and clearly contains no logarithmic divergences for small $m_G^2$. 

The above expression could now be used for studies of general theories: the simplest would be for numerical studies where the potential is evaluated as a function of the expectation values and the derivatives taken numerically, as performed for the MSSM in \cite{Martin:2002iu,Martin:2002wn} and implemented generally in \cite{Goodsell:2014bna}. However, there are potential numerical instabilities when the expectation values of additional scalars are small, and for complicated models many evaluations of the potential are required which can be slow: it is therefore useful to have explicit expressions for the tadpoles, as were derived at two loops in \cite{Goodsell:2015ira}. In the next section we shall compute these for the resummed potential.

\section{Removing infra-red divergences in the minimum condition}
\label{SEC:tadpoles}

In the previous section we derived the resummed two-loop effective potential expanded in $m_G^2$ that explicitly contains no infra-red divergences in its derivatives. In this section we shall present these derivatives. However, we shall also present a new approach to the problem which allows us to calculate the derivatives simply, and so we shall also give our derivations. For the scalar-only diagrams we do this by three methods:
\begin{itemize}
 \item[(i)] The first method is to generalise the approach of \cite{Martin:2014bca,Kumar:2016ltb}, and simply take the derivatives of the resummed potential (\ref{resummed_veff}). However, this has the disadvantage of 
 requiring us to compute the derivative of the rotation matrix elements $\left.\frac{\partial R_{ij}}{\partial\phi_r^0}\right|_{\varphi=v}$ and proves to be cumbersome: there are dramatic simplifications in the final result. 
 \item[(ii)] To avoid the  derivatives of rotation matrix elements, we instead take the derivatives of $\veffr$ before diagonalising the mass matrix and singling  out the Goldstone boson and expanding the potential in $m_G^2$. This leads to a simpler derivation of the results. 
 \item[(iii)] For our third method, we introduce a new approach: we set the Goldstone boson mass ``on-shell'' in the (non-resummed) effective potential. We shall show that this gives the same result as the other methods but (much) more simply, and does not suffer from the problem of needing to exclude Goldstone self interactions by hand. Furthermore, in the next section we shall employ this approach to compute the mass digrams, which would be more complicated using the alternative methods.
\end{itemize}

\subsection{All-scalar diagrams}
\label{SUBSEC:tad_scalar}

\subsubsection{Elimination of the divergences by method (i)}
\label{SUBSUBSEC:scalar_method1}
Generalising the approach of \cite{Kumar:2016ltb} to extract the tadpoles we take the derivatives of equation (\ref{EQ:vtwo_scalar}). Starting with the one-loop potential, we note that, since $m_G^2 + \Delta_G =0$ at the minimum, the derivative of $f (m_G^2 + \Delta_G)$ will vanish. Hence we only require
\begin{align}
\frac{\partial \hat{V}_S^{(1)}}{\partial \phi_r^0} =& \sum_{i \ne G}   \frac{1}{2} A(m_i^2) \lambda^{iik} R_{rk}.
\label{min_cond_1l}
\end{align}
Note that throughout we shall adopt the Einstein convention for summing repeated indices when all indices are to be summed over; when there is an index that is summed over only a subset (i.e. excluding the Goldstone boson indices) we shall write an explicit sum symbol.

For the two-loop terms, recall the scalar part
\begin{align}
\hat{V}_{S}^{(2)} =& \vtwo_{SSS}|_{\rm no\ GB}   +\sum_{j,k\neq G}\frac{1}{4}(\lambda^{Gjk})^2f_{SSS}(0,m_j^2,m_k^2) +\sum_{k\neq G}\frac{1}{4}(\lambda^{GGk})^2f_{SSS}(0,0,m_k^2) \nn\\
& + \vtwo_{SS}|_{m_G^2 = 0} + \frac{1}{2}\Omega_S m_G^2.
\end{align}
Treating each of these pieces in turn we find:
\begin{align}
\frac{\partial\vtwo_{SSS}|_{\rm no\ GB}}{\partial \phi^0_r} =& \sum_{i,j,k \ne G}\bigg[  \frac{1}{4} \lambda^{iil} R_{rl} (\lambda^{ijk})^2 f^{(1,0,0)}_{SSS} (m_i^2, m_i^2; m_j^2, m_k^2)+  \frac{1}{2} \lambda^{ijk} \lambda^{i'jk} (R^T \partial_r R)_{i'i} f_{SSS} (m_i^2, m_j^2, m_k^2) \nn\\
& + \frac{1}{6} \lambda^{ijk} \lambda^{ii' jk} R_{ri'} f_{SSS} (m_i^2, m_j^2, m_k^2) \bigg] \nn\\
=& R_{rl}  \sum_{i,j,k\ne G}\bigg[  \frac{1}{4} \lambda^{ijk} \lambda^{i'jk}\lambda^{i i' l} U_0(m_i^2, m_{i'}^2; m_j^2, m_k^2) - \frac{1}{6} \lambda^{ijk} \lambda^{il jk} I (m_i^2, m_j^2, m_k^2) \bigg]
\end{align}
and similarly we see
\begin{align}
\frac{\partial}{\partial \phi_r^0} \sum_{j,k\neq G}\frac{1}{4}(\lambda^{Gjk})^2f_{SSS}(0,m_j^2,m_k^2) =  R_{rl} \sum_{j,k\neq G} \bigg[& - \frac{1}{2} \lambda^{Gjk} \lambda^{G l jk} I (0, m_j^2, m_k^2) \nn\\
& + \frac{1}{4} \lambda^{Gjk} \lambda^{Gj'k}\lambda^{j j' l} U_0(m_j^2, m_{j'}^2; 0, m_k^2)\bigg],\\
\frac{\partial}{\partial \phi_r^0} \sum_{k\neq G}\frac{1}{4}(\lambda^{GGk})^2f_{SSS}(0,0,m_k^2) =  R_{rl} \sum_{k\neq G} \bigg[& - \frac{1}{2} \lambda^{GGk} \lambda^{GG l k} I (0, 0, m_k^2) \nn\\
& + \frac{1}{4} \lambda^{GGk} \lambda^{GGk'}\lambda^{k k' l} U_0(m_k^2, m_{k'}^2; 0, 0)\bigg].
\end{align}
 Putting this all together we see that they combine to give the compact expression
\begin{align}
\frac{\partial\vtwo_{SSS}|_{\gbm=0}}{\partial \phi_r^0} =& R_{rl}  \sum_{i\ne G,j,k}\bigg[  \sum_{i'}\frac{1}{4} \lambda^{ijk} \lambda^{i'jk}\lambda^{i i' l} U_0(m_i^2, m_{i'}^2; m_j^2, m_k^2)  - \frac{1}{6} \lambda^{ijk} \lambda^{il jk} I (m_i^2, m_j^2, m_k^2) \bigg]\bigg|_{\gbm\rightarrow 0}.
\label{EQ:dVSSS}\end{align}
Next we turn to the $SS$ terms: 
\begin{align}
\frac{\partial\vtwo_{SS}|_{\gbm=0}}{\partial \phi_r^0}=& R_{rl} \sum_{i,j \ne G} \bigg[ - \frac{1}{4} \lambda^{iijj} \lambda^{iil} B(0, m_i^2, m_i^2) A(m_j^2)  +  \frac{1}{2} \lambda^{ii'jj} (R^T \partial_r R)_{i' i} A(m_i^2) A( m_j^2) \bigg] \nn\\
=& \frac{1}{4} R_{rl} \sum_{i,j \ne G}  \lambda^{ii'jj} \lambda^{ii'l}P_{SS} ( m_i^2, m_{i'}^2) A(m_j^2)\big|_{m_G^2 = 0},
\label{EQ:dVSS}\end{align}
where the two terms again combine into a single compact expression. 
The final piece is 
\begin{align}
\frac{1}{2} \Omega_S \frac{\partial m_G^2 }{\partial \phi_r^0} =& \lambda^{GGl} R_{Gl} \sum_{(j,k)\ne (G,G)} \frac{1}{4} (\lambda^{Gjk})^2 R_{SS} (m_j^2, m_k^2)\big|_{m_G^2 = 0} ,
\label{EQ:dVOmegaS}\end{align}
using the expression of $\Omega_S$ from eq.~(\ref{delta1s}). The total scalar tadpole is then the sum of equations (\ref{EQ:dVSSS}), (\ref{EQ:dVSS}) and (\ref{EQ:dVOmegaS}). Clearly the simplicity of the final result compared to the intermediate expressions implies that there should be a simpler way of deriving it -- as indeed we shall show.

\subsubsection{Elimination of the divergences by method (ii)}
\label{SUBSUBSEC:scalar_method2}

From inspection it is clear that the one-loop tadpole is not divergent when we send $m_G^2 \rightarrow 0$.
However, at two loops we found that the process of isolating the divergences in the potential, expanding it in the Goldstone mass, and then taking the derivatives was rather cumbersome due to the derivatives of the mixing matrix elements $R_{ij}$. Instead we could consider taking the derivatives before having cancelled out the divergent parts, and then ensure the cancellations later.  
Hence we rewrite the resummed effective potential as 
\begin{equation}
\label{veffr_vs_veff}
 \veffr=\veff+\frac{1}{16\pi^2}\left(f\big(\gbm+\Delta_G\big)-f(\gbm)\right)-\frac{1}{16\pi^2}\frac{1}{2}A(\gbm)\Delta_G,
\end{equation}
using formulae ~(\ref{exp_2l}) and ~(\ref{resummed_veff}). 
We expect the  terms from the derivative of $-\frac{1}{2}A(\gbm)\Delta_G$ to cancel off the IR divergences in the derivatives of $\veff$.  To show this, we use the expression of $\Delta_G$ derived in eq.~(\ref{delta1s}). 
The relevant contribution to the 
minimum condition at two-loop order is 
\begin{align}
\label{elim_method2}
 &16\pi^2\left.\frac{\partial}{\partial\phi_r^0}\left(-\frac{1}{2}A(\gbm)\Delta_G\right)\right|_{\varphi=v}
 \supset-\frac{1}{2}\left.\frac{\partial\gbm}{\partial\phi_r^0}\right|_{\varphi=v}\llog\gbm (\Delta_1)_S \nn\\
 &=-\frac{1}{2}R_{rp}\lambda^{GGp}\llog\gbm\left(\sum_{(j,k)\neq (G,G)}\frac{1}{2}(\lambda^{Gjk})^2P_{SS}(m_j^2,m_k^2)+\sum_{j\neq G}\frac{1}{2}\lambda^{GGjj}A(m_j^2)\right).
\end{align}

The purely scalar contribution to the \emph{non-resummed} tadpoles is, at one-loop order 
\begin{equation}
 \label{EQ:deriv_1lpart_scalar}
 \left.\frac{\partial\vone_S}{\partial\phi_r^0}\right|_{\varphi=v}=\frac{1}{2} R_{rk}\lambda^{iik} A(m_i^2) 
\end{equation}
and at two loops
\begin{align}
 \label{deriv_2lpart}
 \left.\frac{\partial\vtwos}{\partial\phi_r^0}\right|_{\varphi=v}=R_{rp}&\left(T_{SS}^p+T_{SSS}^p+T_{SSSS}^p\right),
\end{align}
where \cite{Goodsell:2015ira}
\begin{align} 
T_{SS}^p&=\frac{1}{4}\lambda^{jkll}\lambda^{jkp}f^{(1,0)}_{SS}(m_j^2,m_k^2;m_l^2)=\frac{1}{4}\lambda^{jkll}\lambda^{jkp}P_{SS}(m_j^2,m_k^2) A(m_l^2),\\
T_{SSS}^p&=\frac{1}{6}\lambda^{pjkl}\lambda^{jkl}f_{SSS}(m_j^2,m_k^2,m_l^2)=-\frac{1}{6}\lambda^{pjkl}\lambda^{jkl}I(m_j^2,m_k^2,m_l^2),\\
T_{SSSS}^p&=\frac{1}{4}\lambda^{pjj'}\lambda^{jkl}\lambda^{j'kl}f^{(1,0,0)}_{SSS}(m_j^2,m_{j'}^2;m_k^2,m_l^2) = \frac{1}{4}\lambda^{pjj'}\lambda^{jkl}\lambda^{j'kl} U_0(m_j^2,m_{j'}^2;m_k^2,m_l^2) ,
\end{align}
with the notation $f^{(1,0,0)}_{\alpha}$ defined in eq.~(\ref{f_deriv_def}).

In these formulae, we can then consider separately the Goldstone contributions and investigate the divergent terms. We find two types of divergent terms in eq.~(\ref{deriv_2lpart}) :
\begin{itemize}
 \item The first type of divergent term comes from $T_{SS}$, for $j=k=G$, and\footnote{The term with $l=G$ is proportional to $\llog\gbm A(\gbm)$, which tends to zero when $\gbm\rightarrow 0$.} $l\neq G$, and reads
 \begin{equation}
 \label{a*log_div}
  \left.\frac{\partial\vtwos}{\partial\phi_r^0}\right|_{\varphi=v}\supset-\frac{1}{4}R_{rp}\sum_{l\neq G}\lambda^{GGll}\lambda^{GGp}B_0(\gbm,\gbm)A(m_l^2)
  =\frac{1}{4}R_{rp}\sum_{l\neq G}\lambda^{GGll}\lambda^{GGp}\llog\gbm A(m_l^2) 
 \end{equation}
 \item The other divergent terms, coming from $T_{SSSS}^p$ with $j=j'=G$, are
 \begin{equation}
 \label{pss*log_div}
   \left.\frac{\partial\vtwos}{\partial\phi_r^0}\right|_{\varphi=v}\supset\frac{1}{4}R_{rp}\lambda^{pGG}\lambda^{Gkl}\lambda^{Gkl}\llog\gbm P_{SS}(m_k^2,m_l^2)
 \end{equation}
 \item A potentially more dangerous element of those terms, for the particular case $k=l=G$ is not present as $\lambda^{GGG}=0$ (at least up to terms of one-loop order). 
\end{itemize}
All the other terms in $\left.\frac{\partial\vtwo_S}{\partial\phi_r^0}\right|_{\varphi=v}$ are regular in the limit $\gbm\rightarrow 0$.

After relabelling of the indices in the sums, we observe that the $\llog\gbm$ divergences from the terms in eqs.~(\ref{a*log_div}) and~(\ref{pss*log_div}) cancel out perfectly with the ones from eq.~(\ref{elim_method2}). 
We can then take the limit $\gbm\rightarrow 0$ in the one-loop and two-loop parts of the minimum condition: this limit is regular in the one-loop tadpole (\ref{EQ:deriv_1lpart_scalar}) so we recover eq.~(\ref{min_cond_1l}), while we find 
\begin{align}
\label{min_cond_2l}
 \left.\frac{\partial\vtwor_S}{\partial\phi_r^0}\right|_{\varphi=v}&=\frac{1}{4}R_{rp}\left\{\sum_{j,k,l\neq G}\lambda^{jkll}\lambda^{jkp}P_{SS}(m_j^2,m_k^2)A(m_l^2)
 +2\sum_{k,l\neq G}\lambda^{Gkll}\lambda^{Gkp}P_{SS}(0,m_k^2)A(m_l^2)\right\}\nn\\
 &+\frac{1}{6}R_{rp}\lambda^{pjkl}\lambda^{jkl}\left.f_{SSS}(m_j^2,m_k^2,m_l^2)\right|_{\gbm\rightarrow 0}\nn\\
 &+\frac{1}{4}R_{rp}\left\{\sum_{(j,j')\neq(G,G)}\lambda^{pjj'}\lambda^{jkl}\lambda^{j'kl}\left.U_0(m_j^2,m_{j'}^2,m_k^2,m_l^2)\right|_{\gbm\rightarrow 0}\right.\nn\\
 &\quad\quad\quad\quad\quad\quad+\left.\sum_{(k,l)\neq(G,G)}\lambda^{pGG}(\lambda^{Gkl})^2\left.R_{SS}(m_k^2,m_l^2)\right|_{\gbm\rightarrow 0}\right\},
\end{align}
at two-loop order. It is important to notice that all three functions $f_{SSS},\,U_0$ and $R_{SS}$ are regular when one of their arguments goes to zero, hence 
the result we find is indeed free of infrared divergences.

\subsubsection{Elimination of the divergences by setting the Goldstone boson on-shell}
\label{SUBSUBSEC:scalar_method3}

Here we shall introduce a new approach to the Golstone Boson Catastrophe: we shall treat the Goldstone boson mass as an on-shell parameter and enforce that it is identically zero. 
 This means replacing the dimensionally regularised (\drbar or \msbar) Goldstone mass by the on-shell (or pole) mass in the following way
 \begin{equation}
  (\gbm)^{\dimreg}\equiv(\gbm)^\text{OS}-\Pi_{GG}^{(1)}\big((\gbm)^\text{OS}\big) = -\Pi_{GG}^{(1)}\big(0\big) 
\label{EQ:PoleGmass} \end{equation}
where the pole mass is $(\gbm)^\text{OS}=0$. Note that we only need the one-loop relation here, so any mixing in the mass terms between the Goldstone boson and other (pseudo-)scalars is irrelevant -- it would be proportional to $(\Pi_{iG}^{(1)})^2$ and thus a two-loop effect.
When we write the effective potential in terms of the on-shell Goldstone boson mass we should find that it is free of divergences. To do this, we shall start from the dimensionally regularised potential and substitute the Goldstone boson mass in equation (\ref{EQ:PoleGmass}), expanding out to the appropriate loop order; this gives the result that we would obtain by performing the calculation using the on-shell mass with the appropriate counterterms. For our case, we only need to use the one-loop self-energy in the one-loop tadpole; the scalar contribution to the Goldstone boson self-energy at one-loop order is
\begin{equation}
\label{goldstone_1l_selfenergy}
 \Pi_{GG}^{(1), S}\big(p^2\big)=\frac{1}{2}\lambda^{GGjj}A(m_j^2)-\frac{1}{2}(\lambda^{Gjk})^2B(p^2,m_j^2,m_k^2)
\end{equation}
where we again require the result $\lambda^{GGG} = 0$ to leading order -- although in this case we could (if desired) make it an on-shell condition. Applying the above relation to the tadpole in eq.~(\ref{EQ:deriv_1lpart_scalar}) we obtain the following shift to the two-loop tadpole:
\begin{align}
 \frac{1}{2}R_{rp}\lambda^{GGp}A(\gbm)=&\frac{1}{2}R_{rp}\lambda^{GGp}A((\gbm)^\text{OS})-  \frac{1}{2}R_{rp}\lambda^{GGp}\llog(\gbm)^\text{OS}\Pi_{GG}^{(1)}((\gbm)^\text{OS})+ \mathcal{O}(\mathrm{3-loop}) \\
\rightarrow \frac{\partial V^{(2)}_S}{\partial \phi_r^0} ((\gbm)^\text{OS}) &= \left.\frac{\partial V^{(2)}_S}{\partial \phi_r^0}\right|_{m_G^2 \rightarrow (\gbm)^\text{OS}}-\frac{1}{4}R_{rp}\lambda^{GGp}\llog(\gbm)^\text{OS}\left(\lambda^{GGjj}A(m_j^2)-(\lambda^{Gjk})^2B(0,m_j^2,m_k^2)\right). \nn
\end{align}
Since $B(0,m_j^2,m_k^2)=-P_{SS}(m_j^2,m_k^2)$, these shifts correspond exactly to the divergent terms we saw in equations (\ref{a*log_div}) and (\ref{pss*log_div}) and so when we formally take the limit $(\gbm)^\text{OS}\rightarrow 0$ we find exactly the same tadpole given explicitly in (\ref{min_cond_2l}) that we found by the two other methods. This derivation is certainly much faster than the first method, but note that the principle is different to the previous calculations: there is no ad-hoc resummation, nor are we required to expand the potential as a series in $\gbm$. However, perhaps remarkably, we find exactly the same result for the tadpole that remains, implying that, at least at two loops, the two approaches are equivalent. This new approach will prove to be simpler than both previous methods when we turn our attention to mass diagrams; for now we shall simply complete the set of tadpole equations. 

Before moving on to diagrams with fermions, we shall comment on the prescription to follow when there is more than one Goldstone boson. In that case, since the Goldstone bosons are all degenerate the mutual mixing between them becomes a leading-order effect and we must diagonalise the self-energies $\Pi_{GG'}$ on the subspace of indices $G, G'$ which run over all Goldstones. However, we can also easily write this in the non-diagonalised basis as a generalisation of (\ref{EQ:PoleGmass}):
\begin{equation}
  (m_{GG'}^2)^{\dimreg}\equiv(m_G^2)^\text{OS}-\Pi_{GG'}^{(1)}\big((\gbm)^\text{OS}\big) = -\Pi_{GG'}^{(1)}\big(0\big),
\end{equation}
where formally all Goldstone bosons have the same mass $m_G^2$ which we set to zero. Then we can rewrite the tadpole as 
\begin{align}
\frac{1}{2}R_{rp}\lambda^{GG'p}A(m_{GG'}^2)=& \sum_{G} \frac{1}{2}R_{rp}\lambda^{GGp}A((\gbm)^\text{OS})- \sum_{G,G'} \frac{1}{2}R_{rp}\lambda^{GG'p}\llog(\gbm)^\text{OS}\Pi_{GG'}^{(1)}((\gbm)^\text{OS})+ \mathcal{O}(\mathrm{3-loop}) .
\end{align}
If the gauge group of the model of interest is just that of the Standard Model, then clearly the charged and neutral Goldstone bosons cannot mix, so this becomes trivial -- hence in the following we shall restrict for clarity to the one-Goldstone case. However, we shall later write the full result in the general case.

\subsection{Diagrams with scalars and fermions}
\label{SUBSEC:tad_fermion}

The one-loop tadpoles involving fermions are
\begin{align}
\label{deriv_1lpart_fermion}
 \left.\frac{\partial\vone_F}{\partial\phi_r^0}\right|_{\varphi=v} =& R_{rp} T_F^{p}  = -R_{rp} \mathrm{Re} [y^{KLp} M_{KL}] \big(A(m_K^2) + A(m_L^2)\big)  
\end{align}
and these do not present any divergence in the limit of vanishing Goldstone boson mass. The two-loop contributions are \cite{Goodsell:2015ira}, 
\begin{align}
\label{deriv_2lpart_fermion}
 \left.\frac{\partial\vtwofs}{\partial\phi_r^0}\right|_{\varphi=v}=R_{rp}&\left(T_{SSFF}^p+T_{FFFS}^p\right),
\end{align}
where
\begin{align}
\label{expr_Tssff}
 T_{SSFF}^p=&\frac{1}{2}y^{IJk}y_{IJl}\lambda^{klp}f^{(0,0,1)}_{FFS}(m_I^2,m_J^2;m_k^2,m_l^2)\nn\\
 &\quad-\text{Re}\bigg[y^{IJk}y^{I'J'k}M^*_{II'}M^*_{JJ'}\bigg]\lambda^{klp}U_0(m_k^2,m_l^2,m_I^2,m_J^2),\\
\label{expr_Tfffs} 
 T_{FFFS}^p=&2\text{Re}[y^{IJp}y_{IKn}y^{KLn}M^*_{JL}]T_{F\ov{F}FS}(m_I^2,m_J^2,m_K^2,m_n^2)\nn\\
 &\quad+2\text{Re}[y_{IJp}y^{IKn}y^{JLn}M^*_{KL}]T_{FF\ov{F}S}(m_I^2,m_J^2,m_K^2,m_n^2)\nn\\
 &\quad-2\text{Re}[y^{IJp}y^{KLn}y^{MPn}M^*_{IK}M^*_{JM}M^*_{LP}]T_{\ov{F}\ov{F}\ov{F}S}(m_I^2,m_J^2,m_L^2,m_n^2),
\end{align}
with the loop functions from eq.~(II.38) of \cite{Goodsell:2015ira}.

The second term $T_{FFFS}$ is regular when $\gbm\rightarrow 0$, because the loop functions, $B_0$, $I$, $U_0$,
that appear in its expression are all regular when only one of their argument goes to zero. 
However, the $k=l=G$ terms in $T_{SSFF}$ are divergent:
\begin{align}
\label{tadpole_fermpart_div}
 T_{SSFF}^p\supset&\frac{1}{2}y^{IJG}y_{IJG}\lambda^{GGp}\llog\gbm P_{FF}(m_I^2,m_J^2)\nn\\
 &\quad+\frac{1}{2}\text{Re}\bigg[y^{IJG}y^{I'J'G}M^*_{II'}M^*_{JJ'}\bigg]\lambda^{GGp}\llog\gbm P_{\ov{F}\ov{F}}(m_I^2,m_J^2).
\end{align}
After either resummation or setting the Goldstone boson on-shell we find the total, finite, two-loop contribution $T_{SSFF}^p$ in equation (\ref{EQ:TadpoleFermion}) 
and note that $T_{FFFS}^p$ is not modified from eq.~(\ref{expr_Tfffs}).

\subsection{Diagrams with scalars and gauge bosons}
\label{SUBSEC:tad_gauge}

The one-loop tadpole involving (massive) gauge bosons is
\begin{align}
\label{deriv_1lpart_vectors}
 \left.\frac{\partial\vone_{V}}{\partial\phi_r^0}\right|_{\varphi=v} =& R_{rp} T_V^{p}  = \frac{1}{2} R_{rp} g^{aap} A(m_a^2), 
\end{align}
which contains no scalar propagators so has no divergences in the Goldstone boson mass. 

However, the gauge boson contribution to the one-loop scalar self-energy in Landau gauge is \cite{Martin:2003it}:
\begin{align}
\Pi_{ij}^{(1,V)} =& g^{aik} g^{ajk} B_{SV} (m_k^2, m_a^2) + \frac{1}{2} g^{aaij} A_V (m_a^2) + \frac{1}{2} g^{abi} g^{abj} B_{VV} (m_a^2,m_b^2),
\end{align}
where the loop functions are given in \cite{Martin:2003qz,Martin:2003it} but simplify for zero momentum in Landau gauge to
\begin{align}
\left.B_{SV} ( x,y)\right|_{p^2=0} =& 0,  \nn\\
\left.A_{V} ( x)\right|_{p^2=0} =& 3 A(x) + 2 x \delta_{\ov{\rm MS}},  \nn\\
\left. B_{VV} (x,y) \right|_{p^2=0} =& 3 P_{SS} (x,y) + 2\delta_{\ov{\rm MS}} .
\end{align}
Recall that there are six scalar-gauge boson contributions to the two-loop tadpole \cite{Goodsell:2015ira}:
\begin{align}
T_{SSV}^p =&  \frac{1}{2} g^{aij} g^{akj} \lambda^{ikp} f_{SSV}^{(1,0,0)} (m_i^2,m_k^2; m_j^2, m_a^2) + \frac{1}{4}  g^{aij} g^{bij} g^{abp} f_{SSV}^{(0,0,1)} (m_i^2, m_j^2; m_a^2,m_b^2)\\
T_{VS}^p =& \frac{1}{4} g^{abii} g^{abp}  f_{VS}^{(1,0)} (m^2_a, m^2_b;m_i^2) + \frac{1}{4} g^{aaik} \lambda^{ikp}  f_{VS}^{(0,1)} (m^2_a; m^2_i,m_k^2) \\
T_{VVS}^p =& \frac{1}{2} g^{abi} g^{cbi} g^{acp} f_{VVS}^{(1,0,0)}( m_a^2, m_c^2; m_b^2,m_i^2) + \frac{1}{4} g^{abi} g^{abj} \lambda^{ijp}f_{VVS}^{(0,0,1)} (m_a^2, m_b^2;m^2_i,m_j^2).
\end{align}
Of these only three are potentially singular -- $f_{SSV}^{(1,0,0)}, f_{VS}^{(0,1)} $ and $f_{VVS}^{(0,0,1)}  $; from shifting the tadpoles we obtain
\begin{align}
\Delta T^p_{SV} =& -\frac{1}{2} \lambda^{GGr} \blog m_G^2 \bigg[ g^{aGk} g^{aGk} B_{SV} (m_k^2, m_a^2)  + \frac{1}{2} g^{aaGG} A_V (m_a^2) + \frac{1}{2} g^{abG} g^{abG} B_{VV} (m_a^2,m_b^2) \bigg] \nn\\
\equiv& \lambda^{GGr} g^{aGk} g^{aGk} \Delta f_{SSV}^{(1,0,0)} (m_G^2,m_G^2;m_k^2,m_a^2) \nn\\
&+ \lambda^{GGr}  g^{aaGG} \Delta f_{VS}^{(0,1)} (m_a^2,m_G^2,m_G^2)+ \lambda^{GGr}  g^{abG} g^{abG} \Delta f_{VVS}^{(0,0,1)} (m_a^2, m_b^2; m_G^2, m_G^2) 
\end{align}
i.e. they correspond exactly to the potentially singular terms. However, note that $B_{SV}$ term is zero -- and indeed we find that $f_{SSV}^{(1,0,0)} (m_G^2,m_G^2;m_k^2,m_a^2) $ is non-singular; we find
\begin{align}
f_{SSV}^{(1,0,0)} (m_G^2,m_G^2;x,y) =& - R_{SV} (x,y) + \mathcal{O}(m_G^2)\nn\\
f_{VS}^{(0,1)} (x,m_G^2,m_G^2) =& (3 A(x) + 2 x \delta_{\ov{MS}}) \blog m_G^2 + \mathcal{O}(m_G^2)\nn\\
f_{VVS}^{(0,0,1)} (y, z; m_G^2, m_G^2) =& -\big(3 P_{SS} (y,z) + \delta_{\ov{\rm MS}}\big)\blog m_G^2 - R_{VV} (y,z) + \mathcal{O}(m_G^2).
\end{align}
We give the final finite tadpoles in equation (\ref{EQ:TadpoleVector}).

\subsection{Total tadpole}

Here we gather the results of the previous subsections and rewrite them for the most general case, that of multiple Goldstone bosons. 
The total tadpole, after curing the Goldstone boson catastrophe and taking $\gbm \rightarrow 0$ everywhere, is 
\begin{equation}
 \frac{\partial \vtwor}{\partial \phi_r^0} =R_{rp}\bigg[\ov{T}^p_{SS}+\ov{T}^p_{SSS}+\ov{T}^p_{SSSS}+\ov{T}^p_{SSFF}+\ov{T}^p_{FFFS}+\ov{T}^p_{SSV}+\ov{T}^p_{VS}+\ov{T}^p_{VVS}+\ov{T}^p_{FFV}+\ov{T}^p_{\ov{FF}V}+\ov{T}^p_{\text{gauge}}\bigg].
\end{equation}
The all-scalar diagrams are
\begin{align}
\ov{T}^p_{SS}=&\frac{1}{4}\sum_{j,k,l\neq G}\lambda^{jkll}\lambda^{jkp}P_{SS}(m_j^2,m_k^2)A(m_l^2)
 +\frac{1}{2}\sum_{k,l\neq G}\lambda^{Gkll}\lambda^{Gkp}P_{SS}(0,m_k^2)A(m_l^2),\\
\ov{T}^p_{SSS}=&\frac{1}{6}\lambda^{pjkl}\lambda^{jkl}f_{SSS}(m_j^2,m_k^2,m_l^2)\big|_{m_G^2 \rightarrow 0},\\
\ov{T}^p_{SSSS}=&\frac{1}{4}\sum_{(j,j')\neq(G,G')}\lambda^{pjj'}\lambda^{jkl}\lambda^{j'kl}U_0(m_j^2,m_{j'}^2,m_k^2,m_l^2)\nn\\
 &+\frac{1}{4}\sum_{(k,l)\neq(G,G')}\lambda^{pGG'}\lambda^{Gkl}\lambda^{G'kl}R_{SS}(m_k^2,m_l^2),
\end{align}  
where by $(j,j')\neq(G,G') $ we mean that $j, j'$ are not both Goldstone indices.
The fermion-scalar diagrams are
\begin{align}
\label{EQ:TadpoleFermion}
\ov{T}^p_{SSFF}=&\sum_{(k,l)\neq(G,G')}\left\{\frac{1}{2}y^{IJk}y_{IJl}\lambda^{klp}f^{(0,0,1)}_{FFS}(m_I^2,m_J^2;m_k^2,m_l^2)\right. \left.-\text{Re}\bigg[y^{IJk}y^{I'J'k}M^*_{II'}M^*_{JJ'}\bigg]\lambda^{klp}U_0(m_k^2,m_l^2,m_I^2,m_J^2)\right\}\nn\\
 &+\frac{1}{2}\lambda^{GG'p}y^{IJG}y_{IJG'}\left(-I(m_I^2,m_J^2,0)-(m_I^2+m_J^2)R_{SS}(m_I^2,m_J^2)\right)\nn\\
 &-\lambda^{GG'p}\text{Re}\bigg[y^{IJG}y^{I'J'G'}M^*_{II'}M^*_{JJ'}\bigg]R_{SS}(m_I^2,m_J^2), \\
\ov{T}^p_{FFFS} =&T^p_{FFFS}\big|_{\gbm\rightarrow 0},
\end{align}
while the gauge boson-scalar tadpoles are
\begin{align}
 \ov{T}_{SSV}^p =&  T_{SSV}^p \big|_{m_G^2 \rightarrow 0},\nn\\
\ov{T}_{VS}^p =&   {1\over 4} g^{abii} g^{abp}  f_{VS}^{(1,0)} (m^2_a, m^2_b;m_i^2)\big|_{m_G^2 \rightarrow 0} + \sum_{(i,k) \ne (G,G')} {1\over 4} g^{aaik} \lambda^{ikp}  f_{VS}^{(0,1)} (m^2_a; m^2_i,m_k^2), \nn\\
\ov{T}_{VVS}^p =&  \frac{1}{2} g^{abi} g^{cbi} g^{acp} f_{VVS}^{(1,0,0)}( m_a^2, m_c^2; m_b^2,m_i^2)\big|_{m_G^2 \rightarrow 0} + \sum_{(i,j) \ne (G,G')} {1\over 4} g^{abi} g^{abj} \lambda^{ijp}f_{VVS}^{(0,0,1)} (m_a^2, m_b^2;m^2_i,m_j^2) \nn\\
& - {1\over 4} g^{abG} g^{abG'} \lambda^{GG'p} R_{VV} (m_a^2,m_b^2).
\label{EQ:TadpoleVector}\end{align}
Finally the gauge boson-fermion and gauge diagrams are not affected by the Goldstone boson catastrophe, as scalar masses do not appear in them, and can be 
found in the appendix C.2 of \cite{Goodsell:2015ira} 
\begin{align}
 \ov{T}_{FFV}^p=&2g^{aJ}_I\ov{g}^{K}_{bJ}\text{Re}[M_{KI'}y^{I'Ip}]f^{(1,0,0)}_{FFV}(m^2_I,m^2_K;m^2_J,m^2_a)+\frac{1}{2}g^{aJ}_I\ov{g}^{I}_{bJ}g^{abp}f^{(0,0,1)}_{FFV}(m^2_I,m^2_J;m^2_a,m^2_b),\\
 \ov{T}_{\ov{FF}V}^p=&g^{aJ}_Ig^{aJ'}_{I'}\text{Re}[y^{II'p}M^*_{JJ'}]\big[f_{\ov{FF}V}(m^2_I,m^2_J,m^2_a)+M_I^2f_{\ov{FF}V}^{(1,0,0)}(m^2_I,m^2_{I'};m^2_J,m^2_a)\big]\nn\\
 &+g^{aJ}_Ig^{aJ'}_{I'}\text{Re}[M^{IK'}M^{KI'}M^*_{JJ'}y_{KK'p}]f_{\ov{FF}V}^{(1,0,0)}(m^2_I,m^2_{I'};m^2_J,m^2_a)\nn\\
 &+\frac{1}{2}g^{aJ}_Ig^{bJ'}_{I'}g^{abp}M^{II'}M^*_{JJ'}f_{\ov{FF}V}^{(0,0,1)}(m^2_I,m^2_J;m^2_a,m^2_b),\\
 \ov{T}_\text{gauge}^p=&\frac{1}{4}g^{abc}g^{dbc}g^{adp}f^{(1,0,0)}_\text{gauge}(m^2_a,m^2_d;m^2_b,m^2_c).
\end{align}

\section{Mass diagrams in the gaugeless limit}
\label{SEC:masses}

As discussed in the introduction, the scalar masses are among the most interesting electroweak precision observables, and their calculation also suffers from the Goldstone Boson Catastrophe. Earlier literature pointed out that the calculation in the effective potential approximation contains more severe divergences that cannot be solved by resummation, and thus the inclusion of the external momentum is necessary. However, we shall find that there are also divergences that are not regulated by external momentum -- and thus both setting the Goldstone boson on-shell \emph{and} external momentum are required to obtain finite, accurate results. 

On the other hand, the effective potential approximation is still useful and has advantages over a full momentum-dependent result, chief among these being simplicity and speed of calculation. In particular, the evaluation of the loop functions at arbitrary external momentum requires the numerical solution of differential equations \cite{Martin:2003qz} which, although implemented in the fast package {\tt TSIL} \cite{Martin:2005qm}, is still much slower than the zero-momentum functions, and when the functions must be repeatedly called can lead to times orders of magnitude longer for complicated models. Hence we shall consider expanding the two-loop self-energies as a series in $s \equiv - p^2$ (for metric signature $(-,+,+,+)$) as
\begin{align}
\Pi_{ij}^{(2)} (s) =& \frac{\blog (-s)}{s} \Pi_{-1\, l,ij}^{(2)} + \frac{1}{s} \Pi_{-1,ij}^{(2)} + \Pi_{l^2,ij}^{(2)}  \blog^2 (-s)+ \Pi_{l,ij}^{(2)}  \blog (-s) +  \Pi_{0,ij}^{(2)} +  \sum_{k=1}^\infty \Pi_{k,ij}^{(2)} \frac{s^k}{k!}\ 
\label{EQ:Pisapprox}\end{align}
and we shall neglect terms of $\mathcal{O}(s)$, giving a ``generalised effective potential'' approximation: for loop functions where the singular terms $\Pi_{-1\, l,ij}^{(2)},\Pi_{-1,ij}^{(2)},\Pi_{l^2,ij}^{(2)}, \Pi_{l,ij}^{(2)}  $ vanish the result is identical to the second derivative of the effective potential. This approximation is particularly good when the mass of the scalars considered is smaller than the scale of other particles that they couple to; but even when they are similar we find that typically the difference is only a few percent. This should then be within other uncertainties in the calculation for most purposes. 

We shall perform our calculations using our procedure of taking the Goldstone boson mass(es) on-shell as before, working in the general case now of allowing multiple Goldstone bosons throughout. We shall make heavy use of the existing expressions for two-loop scalar self energies from \cite{Martin:2003it}; however, these are only available up to second order in the gauge coupling. Hence we shall be restricted to work in the very popular ``gaugeless limit'' where we neglect the gauge couplings of broken gauge groups (including electromagnetism, since  hypercharge and weak $SU(2)$ are both broken so their gauge couplings are neglected). The two-loop self-energy in this limit can be decomposed as follows:
\begin{align}
\Pi_{ij}^{(2)} =&  \Pi^{S}_{ij} + \Pi^{SF(W)}_{ij} + \Pi^{SF_4 (M)}_{ij} + \Pi^{S_2 F_3 (M)}_{ij} + \Pi^{S_3F_2 (V)}_{ij} + \Pi^{SF_4 (V)}_{ij} + \Pi^{SV}_{ij} + \Pi_{ij}^{FV}.
\end{align}
This consists of scalar-only propagators, diagrams with scalar and fermion propagators, diagrams with scalar and vector propagators, and fermions and vectors. We find that $\Pi^{SF_4 (M)}_{ij} $ and  $\Pi^{SF_4 (V)}_{ij} $ are nonsingular as $m_G^2  \rightarrow 0$ and $s \rightarrow 0$, so the relevant formulae in that limit are equations (B.15) and (B.28) of \cite{Goodsell:2015ira}. Furthermore, in the gaugeless limit the Goldstone bosons do not couple to the vectors, so $\Pi^{SV}_{ij}$ and $ \Pi_{ij}^{FV} $ are unchanged from (B.36) and (B.41) of \cite{Goodsell:2015ira}. However, the remaining diagrams require regulation: our new expressions for $\Pi^S_{ij}$ are presented in section \ref{SUBSEC:mass_scalar};  $\Pi^{SF(W)}_{ij},  \Pi^{S_2 F_3 (M)}_{ij}$ and $\Pi^{S_3F_2 (V)}_{ij}$ are derived in section \ref{SUBSEC:mass_fermion}.

\subsection{All-scalar terms}
\label{SUBSEC:mass_scalar}

The two-loop scalar self-energy contribution with only scalar propagators is given by \cite{Martin:2003it}:
\begin{eqnarray} 
&& \Pi^{S}_{ij} \>=\>
\frac{1}{4} \lambda^{ijkl} \lambda^{kmn} \lambda^{lmn}
\propW_{SSSS} (m_k^2, m_l^2, m_m^2, m_n^2)
+ \frac{1}{4} \lambda^{ijkl} \lambda^{klmm} \propX_{SSS}
(m_k^2, m_l^2, m_m^2)
 \\ && \quad
+ \frac{1}{2} \lambda^{ikl} \lambda^{jkm} \lambda^{lmnn} 
\propY_{SSSS} (m_k^2, m_l^2, m_m^2, m_n^2)
+ \frac{1}{4} \lambda^{ikl} \lambda^{jmn} \lambda^{klmn}
\propZ_{SSSS} (m_k^2, m_l^2, m_m^2, m_n^2)
\nonumber \\ && \quad
+ \frac{1}{6} \lambda^{iklm} \lambda^{jklm} \propS_{SSS} 
(m_k^2,m_l^2,m_m^2) 
+\frac{1}{2} \left (
\lambda^{ikl} \lambda^{jkmn} + \lambda^{jkl} \lambda^{ikmn} \right )
\lambda^{lmn} \propU_{SSSS}
(m_k^2, m_l^2, m_m^2, m_n^2) 
\nonumber \\ && \quad
+ \frac{1}{2} \lambda^{ikl} \lambda^{jkm} \lambda^{lnp} \lambda^{mnp} 
\propV_{SSSSS} (m_k^2, m_l^2, m_m^2, m_n^2, m_p^2)
+ \frac{1}{2} \lambda^{ikm} \lambda^{jln} \lambda^{klp} \lambda^{mnp}
\propM_{SSSSS}(m_k^2, m_l^2, m_m^2, m_n^2, m_p^2) . \nn
\end{eqnarray}
The loop integral functions are recalled in (\ref{EQ:TwoLoopMassFunctions}).

When at most one of the propagators is a Goldstone boson, we can set $m_G^2 \rightarrow 0, s\rightarrow 0$ and use the simplified expressions below (B.2) of \cite{Goodsell:2015ira}. However, for cases including more Goldstone bosons we must look for singularities since, in general, only the $S_{SSS}$ term is regular. Furthermore, we can divide the functions into those regulated by the momentum and those that are not. In particular, by inspection we see that for two or more Goldstone bosons $W, X, Y, V$ can be divergent as $m_G^2 \rightarrow 0$, \emph{even for finite momentum}; this means those terms must be regulated by resummation -- or, in our case, by shifts from the one-loop self energy by putting the Goldstone bosons on shell. On the other hand, the terms $U, M$ and $Z$ must be regulated by including finite momentum.

\begin{figure}
\begin{center}
\includegraphics[width=0.7\textwidth]{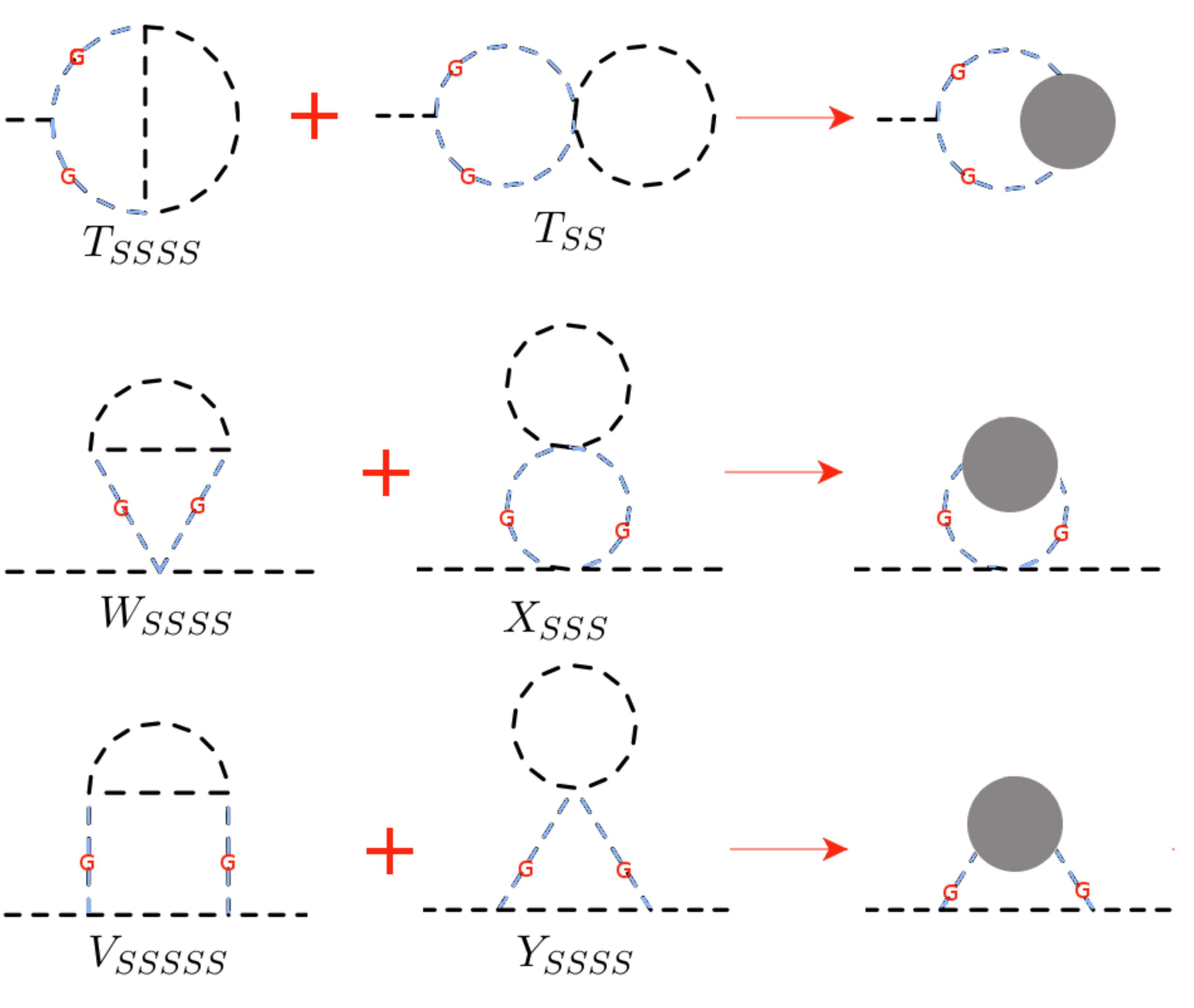}
\end{center}
\caption{Divergent scalar-only diagrams that require regulation (by resummation or using our on-shell scheme), even in the presence of external momentum. The light blue dashed lines marked with a small red ``G'' denote Goldstone boson propagators. The dark blobs in the diagrams on the right-hand side represent full one-loop one-particle-irreducible corrections inserted on the line. On the top line we show the tadpoles (with their clear relation to the sunset and figure-eight diagrams in the potential); on the lower two we show the corrections to the self-energies, which clearly follow the same pattern.}
\label{FIG:Diags}
\end{figure}

It should be noted that the divergences that are not regulated by momentum all involve a Goldstone boson self-energy as a subdiagram. It is then logical to consider how they relate to the divergent terms in the tadpole graphs. If we consider the effective potential approximation and take the derivatives of the tadpoles as in \cite{Goodsell:2015ira}, then we see that the topologies $X,Y,Z$ descend from the $T_{SS}$ graphs; $S,U$ arise from $T_{SSS}$; and $M, V, W$ from $T_{SSSS}$. Then it is clear that, since the $T_{SSS}$ graphs contain no divergences, resummation is irrelevant for $S$ and $U$, while $T_{SS}$ and $T_{SSSS}$ are both divergent when there is part of a Goldstone boson self-energy as a subdiagram. We also see that $W$ and $X$ topologies arise from $T_{SSSS}$ and $T_{SS}$ respectively by replacing a three-point vertex with a four-point one, and likewise $V $ and $Y$ arise by adding a leg connected directly by a propagator to the other leg; we illustrate this whole discussion in figure \ref{FIG:Diags}. Hence we expect that these special divergences should follow the same pattern as the tadpoles, and be cured in the same way. However, we shall also find below some subtleties remain in the $V$ topology. 

\subsubsection{Goldstone shifts}

To determine the effect on the mass diagrams, let us make the shifts using the method of an on-shell Goldstone boson. Recall that the contribution to the one-loop self-energy is
\begin{align}
\Pi_{ij}^{(1),S}(s) =& \frac{1}{2} \lambda^{ijkk} A( m_k^2) - \frac{1}{2} \lambda^{ikl} \lambda^{jkl} B(s,m_k^2, m_l^2) \nn\\
\end{align}
so we can write $\Pi^S_{ij} \rightarrow \Pi_{ij}^S + \Delta \Pi_{ij}^S$ where
\begin{align}
\Delta \Pi_{ij}^S =& - \frac{1}{2} \lambda^{ijGG'} \llog m_G^2 \Pi_{GG'}^{(1),S} (0)+ \lambda^{iGl} \lambda^{jG'l} B^\prime (s,m_G^2, m_l^2) \Pi_{GG'}^{(1),S}(0) \\ 
\equiv& \frac{1}{4} \lambda^{ijGG'} \lambda^{GG'kk} \Delta X_{SSS} (\gbm,\gbm,m^2_k) + \frac{1}{4} \lambda^{ijGG'} \lambda^{Gmn} \lambda^{G'mn} \Delta W_{SSSS} (\gbm,\gbm,m^2_m,m^2_n) \nn\\
& +\frac{1}{2} \lambda^{iGk} \lambda^{jG'k} \lambda^{GG'nn} \Delta Y_{SSSS} (m^2_k,\gbm,\gbm,m^2_n) + \frac{1}{2} \lambda^{ikG} \lambda^{jkG'} \lambda^{Gnp} \lambda^{G'np} \Delta V_{SSSSS} (m^2_k,\gbm,\gbm,m^2_n,m^2_p)\nn
\end{align}
where $B^\prime$ is defined in eq. (\ref{EQ:Bprime_def}), and 
\begin{align}
\Delta X_{SSS} (\gbm,\gbm,m^2_k) \equiv& -A(m^2_k) \llog \gbm = -X_{SSS}(\gbm,\gbm,m^2_k)\nn\\ 
\Delta W_{SSSS} (\gbm,\gbm,m^2_m,m^2_n)\equiv& B(0,m^2_m,m^2_n) \llog \gbm \nn\\ 
\Delta Y_{SSSS} (m^2_k,\gbm,\gbm,m^2_n) \equiv& B^\prime (s,\gbm,m^2_k) A(m^2_n) = -Y_{SSSS}(m^2_k,\gbm,\gbm,m^2_n) \nn\\ 
\Delta V_{SSSSS} (m^2_k,\gbm,\gbm,m^2_n,m^2_p)\equiv& -B^\prime (s,\gbm,m^2_k) B(0,m^2_n,m^2_p) = B^\prime (s,\gbm,m^2_k) P_{SS} (m^2_n,m^2_p).
\end{align}
These \emph{exactly} cancel the divergent parts in the mass diagrams. 
In the case of the $X$ and $Y$ diagrams, they go further and leave no finite parts; for the $W$ diagrams, what remains is
\begin{align}
 W_{SSSS}(\gbm,\gbm,m^2_m,m^2_n)+\Delta W_{SSSS} (\gbm,\gbm,m^2_m,m^2_n)=&U_0(\gbm,\gbm,m^2_m,m^2_n)+\Delta W_{SSSS} (\gbm,\gbm,m^2_m,m^2_n)\nn\\
 =&R_{SS}(m^2_m,m^2_n).
\end{align}
We have no further divergences in $W$ (in particular, $U_0(x,y,0,0)$ is non-singular). 

In the $V$ diagrams there is also a finite piece that remains, since
\begin{align}
 V_{SSSSS} (m^2_k,\gbm,\gbm,m^2_n,m^2_p) =& -  V (m^2_k,\gbm,m^2_n,m^2_p) \nn\\
=& - \overline{V}(m^2_k,m^2_n,m^2_p) + \frac{[P_{SS}(m^2_n,m^2_p) \llog \gbm + R_{SS} (m^2_n,m^2_p) ]}{s-m^2_k} + \mathcal{O}(\gbm) .
\end{align}
Now, using
\begin{align}
B^\prime (s,\gbm,m^2_k) =& -\frac{1}{s-m^2_k} \llog \gbm - \frac{1}{(s-m^2_k)^2} \big[ (m^2_k+s) (B(s,0,m^2_k) - 2) + 2 m^2_k \llog m^2_k\big] + \mathcal{O}(\gbm)
\end{align}
we find that 
\begin{align}
 V_{SSSSS} &(m^2_k,\gbm,\gbm,m^2_n,m^2_p) + \Delta V_{SSSSS} (m^2_k,\gbm,\gbm,m^2_n,m^2_p) = - \overline{V}(m^2_k,m^2_n,m^2_p) \\
&+ \frac{1}{(s-m^2_k)} \bigg[ R_{SS}(m^2_n,m^2_p) + \frac{P_{SS} (m^2_n,m^2_p)}{s-m^2_k} \bigg( (m^2_k+s) (B(s,0,m^2_k) - 2) + 2 m^2_k \llog m^2_k \bigg) \bigg]+ \mathcal{O}(\gbm). \nn
\end{align}
Now we can look at what divergences might remain and need regulating by the momentum. For future reference let us define
\begin{align}
V_{SSSSS} (m^2_k,\gbm,\gbm,m^2_n,m^2_p) + B^\prime (s,\gbm,m^2_k) P_{SS} (m^2_n,m^2_p) \equiv& \tilde{V} (m^2_k,m^2_n,m^2_p). 
\end{align}

 Since we take $\lambda^{GGG} =0$, we never have a divergence from $n=p=G$. On the other hand, when $k=G$ we do have a divergence that needs regulating by the momentum; recalling $B(s,0,0) =-\ov{\log}(-s) + 2$ we can write
\begin{align}
 \tilde{V} (\gbm,m^2_n,m^2_p) =& V_{SSSSS} (\gbm,\gbm,\gbm,m^2_n,m^2_p) + \Delta V_{SSSSS} (\gbm,\gbm,\gbm,m^2_n,m^2_p) \nn\\
=& - \overline{V}(0,m^2_n,m^2_p) + \frac{1}{s} \bigg[ R_{SS}(m^2_n,m^2_p) -P_{SS} (m^2_n,m^2_p)  \ov{\log}( -s) \bigg]+ \mathcal{O}(\gbm). 
\label{EQ:ScalarVG}\end{align}
For the other cases we can set $s=0$ and write 
\begin{align}
 \tilde{V} (m^2_k,m^2_n,m^2_p) =& V_{SSSSS} (m^2_k,\gbm,\gbm,m^2_n,m^2_p) + \Delta V_{SSSSS} (m^2_k,\gbm,\gbm,m^2_n,m^2_p) \nn\\
\overset{k\ne G}{=}& - \overline{V}(m^2_k,m^2_n,m^2_p)+ \frac{1}{k} \bigg[ R_{SS}(m^2_n,m^2_p) - P_{SS} (m^2_n,m^2_p) [ \llog m^2_k - 1]\bigg]+ \mathcal{O}(\gbm). 
\label{EQ:ScalarVk}\end{align}

\subsubsection{Momentum-regulated diagrams}

There are other $V_{SSSSS}$ diagrams that are not regulated by the Goldstone boson shifts. 
 While $V_{SSSSS}(x,y,z,0,0)$, $V_{SSSSS}(0,x,y,0,z)$, $V_{SSSSS}(0,x,y,0,0)$, $V_{SSSSS}(x,0,y,0,0)$ are all regular, the diagrams $V_{SSSSS}(0,0,x,y,z)$ and $V_{SSSSS}(0,0,x,0,y)$ are divergent, and their expression may be found simply by using those for $U(0,0,x,y)$ and $U(0,0,0,x)$ given in appendix \ref{SEC:momentum}:
\begin{align}
V_{SSSSS}(0,0,x,y,z) =& \frac{1}{x} \big[ U_0 (0,x,y,z) - U(0,0,y,z) \big].
\end{align}
All other $V_{SSSSS}$ diagrams are either regular or vanish due to the prefactor $\lambda^{GGG}$.

The remining functions $U_{SSSS}, M_{SSSSS}$ and $Z_{SSSS}$ require regulation by momentum: we give expressions for the expansion of these in appendix \ref{SEC:momentum}.

\subsection{Fermion-scalar diagrams}
\label{SUBSEC:mass_fermion}

The potentially singular mass diagrams are $\Pi_{ij}^{SF(W)}, \Pi_{ij}^{S_2 F_3 (M)} $ and $\Pi_{ij}^{S_3, F_2(V)},$ but among these there are only a subset once more that are regulated by the Goldstone boson shifts; indeed, as in the purely scalar case we find that the topology $M$ is purely regulated by momentum for which all of the limits of the loop functions are provided in appendix  \ref{SEC:momentum}. For the other two, there are exactly four diagrams to regulate, which will match exactly. They have the form \cite{Martin:2003it}:
\begin{align}
\Pi^{SF(W)}_{ij} =& \frac{1}{2} \lambda^{ijkl}
\mathrm{Re} \big[ y^{MNk} y^{M'N'l} M_{MM'} M_{NN'} \big]
\propW_{SS\Fbar\Fbar} (m_k^2, m_l^2, m_M^2, m_N^2)
\nonumber \\ &
+\frac{1}{2} \lambda^{ijkl} y^{MNk} y_{MNl} \propW_{SSFF} (m_k^2, m_l^2, m_M^2, m_N^2), \\
\Pi^{S_3F_2 (V)}_{ij} =& \lambda^{ikl} \lambda^{jkm} \Bigl ( \re \bigl [ y^{NPl} y^{N'P'm} M_{NN'} M_{PP'} \bigr ] \propV_{SSS\Fbar\Fbar} (m_k^2, m_l^2, m_m^2, m_N^2, m_P^2)
\nonumber \\ 
& +\re \bigl [y^{NPl} y_{NPm} \bigr ]\propV_{SSSFF} (m_k^2, m_l^2, m_m^2, m_N^2, m_P^2)\Bigr ),
\end{align}  
and the loop functions are defined in section~\ref{SUBSUBSEC:loopfndef_2l}. \\
As in the scalar case, we look at the shift in the one-loop scalar mass contribution involving Goldstone bosons:
\begin{align}
\Delta \Pi_{ij}^{SF} =& \bigg[ -\frac{1}{2} \lambda^{ijGG'} \llog m_G^2 + \lambda^{iGl} \lambda^{jG'l} B^\prime (s,m_G^2, m_l^2) \bigg] \nn\\
&\quad\quad\quad\quad\times\bigg[\mathrm{Re}(y^{KLG} y_{KLG'}) \Pi_{FF} (m_K^2, m_L^2) + 2\mathrm{Re} (y^{KLG} y^{K'L'G'} M_{KK'} M_{LL'}) \Pi_{\ov{FF}}  \bigg] \nn\\
\equiv & \lambda^{ijGG'} \mathrm{Re}(y^{KLG} y_{KLG'}) \Delta W_{SSFF} +  \lambda^{ijGG'} \mathrm{Re} (y^{KLG} y^{K'L'G'} M_{KK'} M_{LL'})\Delta W_{SS\ov{FF}} \nn\\
& +\lambda^{iGl} \lambda^{jG'l} \mathrm{Re}(y^{KLG} y_{KLG'}) \Delta V_{SSSFF} + \lambda^{iGl} \lambda^{jG'l} \mathrm{Re} (y^{KLG} y^{K'L'G'} M_{KK'} M_{LL'}) \Delta V_{SSS\ov{FF}}
\end{align}
where
\begin{align}
\Pi_{FF} (x,y)\equiv& - \big[(x+y ) P_{SS}(x, y) + A(x) + A(y) \big], \nn\\
\Pi_{\ov{FF}}(x,y) \equiv& - P_{SS} (x, y), 
\end{align}
and compare to the relevant expressions for the loop functions:
\begin{align}
W_{SS\ov{FF}} (m_G^2, m_G^2, x,y) =& -2 U_0(m_G^2,m_G^2,x,y), \\
W_{SSFF} (m_G^2, m_G^2, x,y)= & - (x+ y - m_G^2) U_0(m_G^2,m_G^2,x,y) - I(0,x,y) - \llog \gbm  (A(x) + A(y)), \nn\\
V_{SSSFF} (k,m_G^2,m_G^2,x,y) =& - 2 V_{SSSSS} (k,m_G^2,m_G^2,x,y), \nn\\
V_{SSS\ov{FF}} (k,m_G^2,m_G^2,x,y) =& - (x+ y - m_G^2) V_{SSSSS} (k,m_G^2,m_G^2,x,y) + U(k,m_G^2,x,y) + B^\prime (s,\gbm,k) (A(x) + A(y)). \nn
\end{align}

We should deal with each of these in turn. Firstly for the $W$ topology:
\begin{align}
W_{SSFF} (m_G^2,m_G^2,x,y) + \Delta W_{SSFF} (m_G^2,m_G^2,x,y) \rightarrow& - I(0,x,y) - (x+y) R_{SS} (x,y) , \\
W_{SS\ov{FF}} (m_G^2,m_G^2,x,y) + \Delta W_{SS\ov{FF}} (m_G^2,m_G^2,x,y) \rightarrow& - 2 R_{SS} (x,y) .
\end{align}

For topology $V$, the first combination is proportional to the scalar case in equations (\ref{EQ:ScalarVG}) and (\ref{EQ:ScalarVk}):
\begin{align}
V_{SSS\ov{FF}} (k,m_G^2,m_G^2,x,y) + \Delta V_{SSS\ov{FF}} (k,m_G^2,m_G^2,x,y) =& - 2 V_{SSSSS} (k,m_G^2,m_G^2,x,y) -2 B^\prime (s,m_G^2, k) P_{SS} (x,y) \nn\\
\rightarrow& -2 \tilde{V} (k,x,y),
\end{align}
while the second also contains an additional $U$ function:
\begin{align}
V_{SSSFF} (k,m_G^2,m_G^2,x,y) + \Delta V_{SSSFF} (k,m_G^2,m_G^2,x,y) \rightarrow& - (x+y) \tilde{V} (k,x,y) +  U (k, m_G^2, x,y).
\end{align}
For this case, when $k \ne m_G^2$ it is non-singular as in the scalar case, and when $k = m_G^2$ we require the expansions with finite $s$ from equation  (\ref{EQ:ScalarVG}) and for $U(0,0,x,y)$ from appendix \ref{SEC:momentum}.

\section{Self-consistent solution of the tadpole equations}
\label{SEC:reexpansion}

We have shown how to avoid the Goldstone Boson Catastrophe in general renormalisable field theories, and how this can be applied to calculating neutral scalar masses in the gaugeless limit in a generalised effective potential approximation. However, as they have been formulated the tadpole equations still require an iterative solution, because the same masses appear in the loop functions as we are solving for: recall in equation (\ref{EQ:tadpolesolution}) that $m_{ij}^2 = - \delta_i \delta_{ij} + \hat{m}_{0,ij}^2$ but the $m_{ij}^2$ appear in the $\delta_i$. We cannot do the same as we did for the Goldstone boson and put the other scalars on shell; however, we can follow \cite{Kumar:2016ltb} and use (\ref{EQ:tadpolesolution}) to re-expand the masses to one-loop order in the one-loop tadpole,  then we use the tree-level masses in the loop functions and solve the tadpole equations perturbatively instead of iteratively as described in the introduction.  Let us define a set of masses $\bar{m}^2 = \{m_G^2, \tilde{m}_{i \ne G}^2\}$ i.e. we use the on-shell mass for the Goldstone, and the tree-level masses for the other scalars. To single out the Goldstone boson we use the tree-level mixing matrix $\tilde{R}_{kG}$ which in any case should correspond to the all-loop expression, depending as it does only on the symmetries and vevs. Then we can define the pertubation to the tree-level mass-matrix to be
\begin{align}
\Delta_{ij} \equiv& - \Rt_{ki} \Rt_{kj} \delta_k 
\end{align} 
and we can expand as usual in perturbation theory using $\Delta_{ij} = \frac{1}{16\pi^2}\Delta_{ij}^{(1)} + \frac{1}{(16\pi^2)^2}\Delta_{ij}^{(2)} + ...$ to find that we should shift the tadpoles according to
\begin{equation}
 \frac{\partial \vtwor}{\partial \phi_r^0}(m^2) = \frac{\partial \vtwor}{\partial \phi_r^0}(\bar{m}^2)+\frac{1}{2} \sum_{(i, i')\ne (G,G')} \Rt_{rl} \tilde{\lambda}^{ii'l} \Delta_{i i'}^{(1)} P_{SS} (\bar{m}_i^2, \bar{m}_{i'}^2).
\end{equation}
By $(i, i')\ne (G,G') $ we mean that the sum over $(i,i')$ excludes the cases where \emph{both} $i$ and $i'$ are Goldstone boson indices. 
This allows us to express the $\delta_i$ entirely in terms of the tree-level $\tilde{m}^2$ parameters and obtain a perturbative expansion for $m^2$ -- note that we should also replace all of the couplings $\lambda^{ijk}, \lambda^{ijkl}$  etc and rotation matrices $R_{ij}$ with their tree-level values $\tilde{\lambda}^{ijk}, \tilde{\lambda}^{ijkl}, \Rt_{ij}$ (we already implicitly used this to disregard the $\lambda^{GGG}$ terms). The only subtlety occurs when $\tilde{m}_i^2 = \tilde{m}_j^2$ for some $i, j$ which is not ensured by a symmetry so that $\Delta_{ii} \ne \Delta_{jj}$; in that case as usual the $\Rt$ matrix must be modified to diagonalise $\Delta_{ij}^{(1)}$ on those indices. However the expression above is still valid in that case. Note that the shift only occurs for scalar propagators in the one-loop diagrams, which is why there is no modification of the fermionic or vector tadpole diagrams.

We can apply the same procedure to use the tree-level masses in the mass diagrams: after some algebra we find (in the gaugeless limit -- otherwise we will have some additional shifts from scalar-vector diagrams) that 
\begin{align}
\Pi^{(2)}_{ij} (s,m^2) = \Pi^{(2)}_{ij} (s,\bar{m}^2) + \sum_{(k,k') \ne (G,G')}\left(\frac{1}{2}\tilde{\lambda}^{ijkk'}\Delta_{kk'}^{(1)}P_{SS}(\bar{m}_{k}^2,\bar{m}_{k'}^2)-\tilde{\lambda}^{ikl}\tilde{\lambda}^{jk'l}\Delta_{kk'}^{(1)}C(s,s,0,\bar{m}_k^2,\bar{m}^2_l,\bar{m}_{k'}^2)\right) \nn
\end{align}
where we used the usual $C$ function defined in eq.~(\ref{C_def}). These together then allow us to determine the scalar masses to be the values of $s$ that give solutions to:
\begin{align}
0 =& \mathrm{Det} \bigg[ s \delta_{ij} - m^2_{0,ij} + \delta_i (\bar{m}^2) \delta_{ij} -  \frac{\Pi^{(1)}_{ij} (s,\bar{m}^2)}{16\pi^2} -  \frac{\Pi^{(2)}_{ij} (s,\bar{m}^2)}{(16\pi^2)^2}  \\
&-\frac{\delta_{ij}}{2} \frac{1}{(16\pi^2)^2} \sum_{(j, j')\ne (G,G')} \Rt_{il} \tilde{\lambda}^{jj'l} \Delta_{j j'}^{(1)} P_{SS} (\bar{m}_j^2, \bar{m}_{j'}^2) \nn\\
& - \frac{\Rt_{ii'}\Rt_{jj'}}{(16\pi^2)^2} \sum_{(k,k') \ne (G,G')}\bigg( \frac{1}{2}\tilde{\lambda}^{i'j'kk'}\Delta_{kk'}^{(1)}P_{SS}(\bar{m}_{k}^2,\bar{m}_{k'}^2)-\tilde{\lambda}^{i'kl}\tilde{\lambda}^{j'k'l}\Delta_{kk'}^{(1)}C(s,s,0,\bar{m}_k^2,\bar{m}^2_l,\bar{m}_{k'}^2) \bigg) \bigg] .\nn
\end{align}
Typically in spectrum generators the two-loop corrections are computed at fixed momentum and then the eigenvalues of the above matrix computed iteratively. Since we have given the expansion of all the loop functions relevant for the two-loop corrections up to terms of order $\mathcal{O}(s)$, this could be generalised to include our simple momentum dependence for the two-loop part as in equation (\ref{EQ:Pisapprox}) without significant loss of speed since the computationally expensive parts of the two-loop functions would only need to be evaluated once. However, since all of the expansions are strictly valid only up to two-loop order, the equation above could be solved perturbatively itself with no significant loss of accuracy.

\section{Conclusions}
\label{SEC:conclusions}

We have presented a solution to the Goldstone Boson Catastrophe in general renormalisable theories to two-loop order. We showed that the approach of Goldstone boson resummation is equivalent (at least at two-loop order) to an on-shell scheme for the Goldstone boson(s), the latter being much more convenient calculationally. We then showed how there are a set of self-energy diagrams that also exhibit the Goldstone Boson Catastrophe even when external momentum is included -- but that our solution naturally avoids those singularities. We were then able to give expressions for a ``generalised effective potential approximation'' for neutral scalar masses in the gaugeless limit, that are free of infra-red divergences and give a good approximation to the full momentum-dependent result. This also included the re-expansion of the masses in terms of the values obtained from the tree-level tadpole equations, allowing a self-consistent solution of the tadpole equations (i.e. equations where no terms to be solved for appear on both left and right hand sides). 

The expressions contained in this paper should now allow simple infra-red safe calculations in a wide variety of theories. Most practically, it would be simple to implement them in a package such as \SARAH, to enable automated calculations for \emph{any} model and avoid the problems seen, for example, in \cite{Goodsell:2014pla,Goodsell:2016udb,Athron:2016fuq}, with the existing implementation. This should also enable more reliable and accurate explorations of the parameter space of many models; in particular for non-supersymmetric models (such as the two-Higgs-doublet model),  where the existing ``solution''\footnote{See \cite{Goodsell:2014bna}, appendix 2b of \cite{Goodsell:2014pla} and especially section 2 of \cite{Goodsell:2016udb} for a description of the approach in \SARAH.} to the Goldstone Boson Catastrophe  is not particularly successful, relying as it does on there being a gauge-coupling dependent part of the scalar potential (as in supersymmetric theories).

However, it would also be interesting to explore further many aspects of the problem more generally: the two-loop mass-diagram calculation to quartic order in the gauge couplings; the link between resummation and our on-shell scheme; and also the extension to higher orders. Indeed, these three topics are linked: in \cite{Elias-Miro:2014pca}, it was shown that the momentum dependence of the self-energy in the resummation was necessary for the resummation of certain subleading divergences. By reorganising the expansion in terms of $\tilde{M}_G^2\equiv m_G^2 + \Pi_g(0),$ they showed that the one-loop resummed potential (\ref{EQ:Veff1}) -- which contributes the most divergent parts -- can be rewritten as
\begin{align}
V_\text{eff}^{(1)} =& -\frac{i}{2} C \int d^d k \bigg( \log ( - k^2 + \tilde{M}_G^2 ) - \sum_{L=2}^\infty \frac{1}{(L-1)} \bigg[ \frac{- [\Pi_g(k^2) - \Pi_g(0)]}{\tilde{M}_G^2 - k^2} \bigg]^{L-1} \bigg) \nn\\
=& \frac{1}{4} f( \tilde{M}_G^2) + \mathcal{O}(\tilde{M}_G^4 \blog \tilde{M}_G^2). \nn
\end{align}
This shows that the momentum dependence cannot contribute a more divergent term than $\mathcal{O}(\tilde{M}_G^4 \blog \tilde{M}_G^2)$. Hence if we rewrite the diagrams in terms of $\tilde{M}_G^2$ -- similar to our on-shell scheme -- then the momentum dependence of the self-energy will disappear from the tadpole condition; indeed, \cite{Martin:2014bca} did not require momentum dependence. This also shows that for higher-order contributions it may be most efficient to perform the calculations directly in such a scheme, rather than work in a pure minimal subtraction scheme and then apply the shifts. On the other hand, it could be relevant for the mass diagrams: \cite{Elias-Miro:2014pca} showed that a term of order $\tilde{M}_G^4 \blog \tilde{M}_G^2 $ in the two-loop potential arising from a diagram with a W-boson, charged Goldstone boson and \emph{photon} could be resummed by including the momentum dependence in the self-energy; the masslessness of the photon giving rise to additional infra-red divergences. This issue did not arise here because we worked up to two-loop order, and in the gaugeless limit for the mass diagrams (so that the photon cannot contribute, and its contribution to the tadpoles is benign). It would certainly be interesting to explore this in the future.

\section*{Acknowledgements}

M.~D.~G.~thanks Florian Staub for collaboration and discussion on related topics. We 
both thank Pietro Slavich for many interesting discussions, and for very detailed comments on the 
first version of this paper.
This work was supported in part by French state funds managed by the
Agence Nationale de la Recherche (ANR), in the context of the LABEX
ILP (ANR-11-IDEX-0004-02, ANR-10-LABX-63), and M.~D.~G. acknowledges
support from the ANR grant ``HiggsAutomator'' (ANR-15-CE31-0002). J.~B.~was supported by a
scholarship from the Fondation CFM.

\appendix

\section{Loop functions}
\label{SEC:loopfn}
Throughout our work, we have followed closely the notations of \cite{Martin:2003qz}, however we present in this appendix the loop functions and the notations 
that were used. 
These definitions of loop functions use Euclidean momentum integrals in dimensional reduction to $d=4-2\epsilon$ dimensions, 
and involve the loop factor
\begin{equation}
 C=16\pi^2\frac{\mu^{2\epsilon}}{(2\pi)^d}.
\label{EQ:definitionC}\end{equation}
We also recall the following shorthand notations
\begin{equation}
\label{llog_def}
 \llog x\equiv\log\frac{x}{Q^2},
\end{equation}
where $Q^2=4\pi e^{-\gamma_E}\mu^2$ is the renormalisation scale squared.

\subsection{Definition of loop functions}
\label{SUBSEC:loopfndef}
\subsubsection{One-loop functions}
In the expression of the one-loop effective potential, we make use of the function $f$ defined as
\begin{equation}
\label{expr_f}
 f(x)\equiv\frac{x^2}{4}\bigg(\llog x-\frac{3}{2}\bigg)
\end{equation}

Two important one-loop functions that will appear in the expression of the effective potential, of its derivatives and in the self-energies are  
the finite parts of
\begin{align}
 \textbf{A}(x)&\equiv C\int\frac{d^dk}{k^2+x}\\
 \textbf{B}(p^2,x,y)&\equiv C\int\frac{d^dk}{(k^2+x)((p-k)^2+y)},
\end{align}
namely
\begin{align}
\label{expr_A}
 A(x)&\equiv\lim_{\epsilon\rightarrow 0}\bigg(\textbf{A}(x)+\frac{x}{\epsilon}\bigg)= x(\llog x-1)=2\frac{d}{dx}f(x),\\
 B(p^2,x,y)&\equiv\lim_{\epsilon\rightarrow 0}\bigg(\textbf{B}(p^2,x,y)-\frac{1}{\epsilon}\bigg)= -\llog p^2-f_B(x_+)-f_B(x_-),
\end{align}
where 
\begin{equation}
 f_B(x)=\log(1-x)-x\log\left(1-\frac{1}{x}\right)-1,
\end{equation}
and 
\begin{equation}
 x_\pm=\frac{p^2+x+y\pm\sqrt{(p^2+x+y)^2-4p^2x}}{2p^2}.
\end{equation}
In two-loop order expressions, the function $J$ is sometimes used, although it is equal to $A$
\begin{equation}
 J(x)=A(x).
\end{equation}
A limit of particular interest of $B$ is the limit of vanishing external momentum, that we denote $B_0$, and is related to the $P_{SS}$ function we have used
\begin{equation}
 B(p^2,x,y)\underset{p^2\rightarrow 0}{\longrightarrow}B_0(x,y)=-P_{SS}(x,y)\equiv-\frac{A(x)-A(y)}{x-y}.
\end{equation}
and furthermore, we have that 
\begin{equation}
 B_0(x,x)=-\llog x \Leftrightarrow P_{SS}(x,x)=\llog x
\end{equation}
The derivative of the $B$ function with respect to one of the mass arguments is also used, with the notation
\begin{equation}
\label{EQ:Bprime_def}
B^\prime(p^2,x,y)=\frac{\partial}{\partial x}B(p^2,x,y).
\end{equation}

For the fermion and gauge boson contributions to the scalar self-energy we also use the functions $P_{FF}$, $P_{\ov{F}\ov{F}}$ and $P_{VV}$ related 
to $A$ and $P_{SS}$ as
\begin{align}
 P_{FF}(x,y)&\equiv-2\,\frac{xA(x)-yA(y)}{x-y}=-A(x)-A(y)-(x+y)P_{SS}(x,y),\\
 P_{\ov{F}\ov{F}}(x,y)&\equiv-2P_{SS}(x,y),\\
 P_{VV}(x,y)&\equiv3P_{SS}(x,y).
\end{align}

In the context of the reexpansion of the mass diagrams, we also make use of the one-loop three-point function $C(p_1^2,p_2^2,(p_1+p_2)^2,x,y,z)$, which is the finite 
part of the following integral
\begin{equation}
\label{C_def}
 \textbf{C}(p_1^2,p_2^2,(p_1+p_2)^2,x,y,z)\equiv -C\int\frac{d^dk}{(k^2+x)((k-p_1)^2+y)((k-p_1-p_2)^2+z)}.
\end{equation}

\subsubsection{Two-loop functions}
\label{SUBSUBSEC:loopfndef_2l}
We recall the definition of the following two-loop integrals
\begin{align}
 \textbf{S}(x,y,z)&\equiv C^2\int d^dk\int d^dq\frac{1}{(k^2+x)(q^2+y)((k+q-p)^2+z)},\\
\label{U_integral} 
 \textbf{U}(x,y,z,u)&\equiv C^2\int d^dk\int d^dq\frac{1}{(k^2+x)((k-p)^2+y)(q^2+z)((k+q-p)^2+u)},\\
\label{M_integral}
 \textbf{M}(x,y,z,u,v)&\equiv C^2\int d^dk\int d^dq\frac{1}{(k^2+x)(q^2+y)((k-p)^2+z)((q-p)^2+u)((k-q)^2+v)}.
\end{align}
of which we take the finite parts
\begin{align}
 S(x,y,z)&=\lim_{\epsilon\rightarrow 0}[\textbf{S}(x,y,z)-(\textbf{A}(x)+\textbf{A}(y)+\textbf{A}(z))/\epsilon-(x+y+z)/2\epsilon^2-(p^2/2-x-y-z)/2\epsilon], \nn\\ 
 U(x,y,z,u)&=\lim_{\epsilon\rightarrow 0}[\textbf{U}(x,y,z,u)-\textbf{B}(p^2,x,y)/\epsilon+1/2\epsilon^2-1/2\epsilon],\label{expr_U0}\\
 U_0(x,y,z,u)&\equiv U(x,y,z,u)|_{p^2=0} =\frac{I(x,z,u)-I(y,z,u)}{y-x},\\
 M(x,y,z,u,v)&=\lim_{\epsilon\rightarrow 0}\textbf{M}(x,y,z,u,v).
\end{align}
We also require the related functions (where $\ov{V}$ differs slightly from \cite{Martin:2003it})
\begin{align}
I(x,y,z)&\equiv S(x,y,z)|_{p^2=0},\label{expr_I} \\
V(x,y,z,u) & \equiv - \frac{\partial}{\partial y} U(x,y,z,u), \\
\ov{V} (x,y,z) &\equiv \lim_{u \rightarrow 0} \bigg[ V(x,u,y,z) - \frac{1}{s-x} \frac{\partial}{\partial u} I(u,y,z)\bigg].
\end{align}
The integral $I$ is symmetric on all three indices, and thus $U_0$ is symmetric on $x \leftrightarrow y$ and $z \leftrightarrow u$ separately etc; the $I$ integral is  fundamental for the two-loop effective potential, all other functions being obtained from it and $A(x)$. It can be written explicitly although the expression is rather involved; it can be found in equations (D1) to (D3) of \cite{Degrassi:2009yq} although it was first derived in \cite{Ford:1992pn}. Here we note the useful limiting cases
\begin{align}
I(x,y,0) =& \frac{1}{2}\bigg(-5x - 5y + (-x + y)\llog^2 x + 4y\llog y +   \llog x (4x - 2y\llog y)-2(x - y)\mathrm{Li}_2 (1 - y/x)\bigg), \nn\\
I(x,x,x) =& \frac{3}{2} x(-5 + 4 \llog x - \llog^2 x + c_{xxx}), \nn\\
I(x,x,0) =& x(-5 + 4 \llog x - \llog^2 x), \nn\\
I(x,0,0) =& - x \bigg(\frac{1}{2}  \llog^2x + 2 \llog x - \frac{5}{2}- \frac{\pi^2}{6} \bigg),
\end{align}
where $c_{xxx}\approx 2.3439$ is a constant.

The two-loop functions appearing in the effective potential were defined in \cite{Martin:2001vx} and read
\begin{align}
 f_{SSS}(x,y,z)&=-I(x,y,z),\\
 f_{SS}(x,y)&=J(x,y),\\
 f_{FFS}(x,y,z)&=J(x,y)-J(x,z)-J(y,z)+(x+y-z)I(x,y,z),\\
 f_{\ov{F}\ov{F}S}(x,y,z)&=2I(x,y,z),\\
 f_{SSV}(x,y,z)&=\frac{1}{z}\bigg[(-x^2-y^2-z^2+2xy+2xz+2yz)I(x,y,z)+(x-y)^2I(0,x,y)\nn\\
 &\quad\quad\quad+(y-x-z)J(x,z)+(x-y-z)J(y,z)+zJ(x,y)\bigg]\nn\\
 &\quad+2\left(x+y-\frac{z}{3}\right)J(z),
\end{align}
where
\begin{align}
 J(x,y)&\equiv J(x)J(y)=A(x)A(y).
\end{align}
To these functions we must also add the scheme dependent functions $f_{VS}$, $f_{VVS}$, $f_{FFV}$, $f_{\ov{FF}V}$ and $f_\text{gauge}$ 
that we give for the $\drbar$ and \msbar schemes (slightly modifying the notation of \cite{Martin:2001vx})
\begin{align}
 f_{VS}(x,y)=&3J(x,y)+\delta_\msbar 2xJ(y),\\
 f_{VVS}(x,y,z)=&\frac{1}{4xy}\bigg[(-x^2-y^2-z^2-10xy+2xz+2yz)I(x,y,z)\nn\\
 &\quad\quad+(x-z)^2I(0,x,z)+(y-z)^2I(0,y,z)-z^2I(0,0,z)\nn\\
 &\quad\quad+(z-x-y)J(x,y)+yJ(x,z)+xJ(y,z)\bigg]\nn\\
 &+\frac{1}{2}J(x)+\frac{1}{2}J(y)+\delta_\msbar\big(2J(z)-x-y-z\big),\\
 f_{FFV}(x,y,z)=&\frac{1}{z}\bigg[(x^2+y^2-2z^2-2xy+xz+yz)I(x,y,z)-(x-y)^2I(0,x,y)\nn\\
 &\quad+(x-y-2z)J(x,z)+(y-x-2z)J(y,z)+2zJ(x,y)\bigg]\nn\\
 &+2\left(-x-y+\frac{z}{3}\right)J(z)-\delta_\msbar\big(2xJ(x)+2yJ(y)-(x+y)^2+z^2\big),\\
 f_{\ov{FF}V}(x,y,z)=&6I(x,y,z)+\delta_\msbar\big(2(x+y+z)-4J(x)-4J(y)\big),\\
 f_\text{gauge}(x,y,z)=&\frac{1}{4xyz}\bigg[(-x^4-8x^3y-8x^3z+32x^2yz+18y^2z^2)I(x,y,z)\nn\\
 &\quad\quad\quad+(y-z)^2(y^2+10yz+z^2)I(0,y,z)+x^2(2yz-x^2)I(0,0,x)\nn\\
 &\quad\quad\quad+(x^2-9y^2-9z^2+9xy+9xz+14yz)xJ(y,z)\nn\\
 &\quad\quad\quad+\left(22y+22z-\frac{40}{3}x\right)xyzJ(x)+\delta_\msbar\big(4x^3yz+48xy^2z^2+8x^2yzJ(x)\big)\bigg]\nn\\
 &+(x\leftrightarrow y)+(x\leftrightarrow z)
\end{align}
where 
\begin{eqnarray}
 \delta_\msbar=\left\{\begin{array}{r c l}
                      1&\text{ in the }\msbar\text{ scheme}\\
                      0&\text{ in the }\drbar\text{ scheme}\\
                     \end{array}
\right.
\end{eqnarray}

Taking derivatives of these functions with respect to one argument is required for the two-loop tadpoles, and we use the notations
\begin{align}
\label{f_deriv_def}
 f^{(1,0,0)}_\alpha(x,y;z,u)=\frac{f_\alpha(x,z,u)-f_\alpha(y,z,u)}{x-y}\nn\\
 f^{(0,0,1)}_\alpha(x,y;z,u)=\frac{f_\alpha(x,y,z)-f_\alpha(x,y,u)}{z-u}
\end{align}
For the mass diagrams, we require the following loop integral functions:
\begin{eqnarray}
\propW_{SSSS}(x,y,z,u) &=& [I(x,z,u) - I(y,z,u)]/(y-x),
\\
\propX_{SSS}(x,y,z) &=& J(z) P_{SS}(x,y) ,
\\
\propY_{SSSS}(x,y,z,u) &=& J(u) [B(p^2,x,z) - B(p^2,x,y)]/(y-z),
\\
\propZ_{SSSS}(x,y,z,u) &=& B(p^2,x,y) B(p^2,z,u) ,
\\
\propS_{SSS}(x,y,z) &=& -S(x,y,z) ,
\\
\propU_{SSSS}(x,y,z,u) &=& U(x,y,z,u) ,
\\
\propV_{SSSSS}(x,y,z,u,v) &=& [U(x, y, u, v) -U(x, z, u, v)]/(y - z) ,
\\
\propM_{SSSSS}(x,y,z,u,v) &=& -M(x,y,z,u,v) ,
\\
\propW_{SSFF}(x,y,z,u) &=&\frac{1}{x-y}\left[(z+u-x)I(x,z,u)-A(x)[A(z)+A(u)]\right]+(x\leftrightarrow y),
\\
\propW_{SS\ov{FF}}(x,y,z,u) &=& -2\propW_{SSSS}(x,y,z,u),
\\
\propV_{SSSFF}(x,y,z,u,v) &=& \frac{(y-u-v)U(x,y,u,v)+[A(u)+A(v)]B(s,x,y)}{y-z}+(y\leftrightarrow z),
\\
\propV_{SSS\ov{FF}}(x,y,z,u,v) &=& -2 \propV_{SSSSS}(x,y,z,u,v).
\label{EQ:TwoLoopMassFunctions}\end{eqnarray}

\subsection{Small $\gbm$ expansion}
\label{SUBSEC:smallGexp}

For completeness, we recall equations (3.7)-(3.10) from \cite{Kumar:2016ltb} for the expansion of the loop functions appearing in the two-loop effective 
potential
\begin{align}
\label{fsss_exp}
 f_{SSS}(\gbm,x,y)&=f_{SSS}(0,x,y)+P_{SS}(x,y)A(\gbm)+R_{SS}(x,y)\gbm+\mathcal{O}(m_G^4),\\
\label{exp_fsss_2g} 
 f_{SSS}(\gbm,\gbm,x)&=f_{SSS}(0,0,x)+2P_{SS}(0,x)A(\gbm)+2R_{SS}(0,x)\gbm+\mathcal{O}(m_G^4),\\
 f_{SS}(\gbm,x)&=A(x)A(\gbm),\\
 f_{FFS}(\gbm,x,y)&=f_{FFS}(0,x,y)+P_{FF}(x,y)A(\gbm)+R_{FF}(x,y)\gbm+\mathcal{O}(m_G^4),\\
 f_{\ov{F}\ov{F}S}(\gbm,x,y)&=f_{\ov{F}\ov{F}S}(0,x,y)+P_{\ov{F}\ov{F}}(x,y)A(\gbm)+R_{\ov{F}\ov{F}}(x,y)\gbm+\mathcal{O}(m_G^4),\\
 f_{SSV}(\gbm,x,y)&=f_{SSV}(0,x,y)+R_{SV}(x,y)\gbm+\mathcal{O}(m_G^4),\\
 f_{SSV}(\gbm,\gbm,x)&=f_{SSV}(0,0,x)+2R_{SV}(0,x)\gbm+\mathcal{O}(m_G^4),\\
 f_{VS}(\gbm,x)&=3A(x)A(\gbm)+2x\delta_\msbar A(\gbm),\\
 f_{VVS}(\gbm,x,y)&=f_{VVS}(0,x,y)+(P_{VV}(x,y)+2\delta_\msbar)A(\gbm)+(R_{VV}(x,y)-\delta_\msbar)\gbm+\mathcal{O}(m_G^4),
\end{align}
where the $R$ functions are defined in \cite{Kumar:2016ltb} as 
\begin{align}
 R_{SS}(x,y)=&\{(x+y)^2+2A(x)A(y)-2x A(x)-2y A(y)\nn\\
 &+(x+y)I(0,x,y)\}/(x-y)^2,\\
 R_{FF}(x,y)=&-\bigg[(x+y)\{(x+y)^2+2A(x)A(y)-2x A(x)-2y A(y)+(x+y)^2\}\nn\\
 &+2(x^2+y^2)I(0,x,y)\bigg]/(x-y)^2,\\
 R_{\ov{FF}}(x,y)=&-2R_{SS}(x,y),\\
 R_{SV}(x,y)=&\frac{1}{y}\bigg(3(x+y)I(0,x,y)-3xI(0,0,x)+3A(x)A(y)+2xy+y^2\bigg),\\
 R_{VV}(x,y)=&\frac{1}{4xy(x-y)^2}\bigg[3A(x)A(y)\big(x^2+y^2+6xy\big)-24xy\big(xA(x)+yA(y)\big)\nn\\
 &\quad\quad\quad\quad\quad\quad+14xy(x^2+y^2)+20x^2y^2+3(x+y)^3I(0,x,y)\nn\\
 &\quad\quad\quad\quad\quad\quad-3(x-y)^2\big(xI(0,0,x)+yI(0,0,y)\big)\bigg].
\end{align}

One can see from the expression~(\ref{expr_I}) that $I(=-f_{SSS})$ is regular for any number of its arguments vanishing. 
Using eq.~(\ref{expr_U0}) and~(\ref{fsss_exp}), we can find the expansion of $U_0(x,y,\gbm,\gbm)$
\begin{align}
\label{U0_smallmg_exp}
 U_0 (x,y,\gbm,\gbm) =& - \frac{d}{d\gbm} I(\gbm,x,y) =\frac{d}{d\gbm} f_{SSS}(\gbm,x,y)\nn\\
=& P_{SS} (x,y) \llog \gbm + R_{SS} (x,y) + ...
\end{align}
For the derivatives of the two-loop $f$ functions, we use the following expansions
\begin{align}
f_{FFS}^{(0,0,1)} (x, y, \gbm, \gbm)=& - \llog\gbm [ J(x) + J(y)] - I(x, y, 0) - (x + y) U_0(x, y, \gbm,\gbm) \\ 
=& -\llog\gbm \bigg[ (x+y)P_{SS} (x,y) + A(x) + A(y) \bigg]  - I(x,y,0) - (x+y) R_{SS} (x,y)+ \mathcal{O}(m_G^2),\nn\\
f_{SSV}^{(1,0,0)} (m_G^2,m_G^2;x,y) =& - R_{SV} (x,y) + \mathcal{O}(m_G^2)\\
f_{VS}^{(0,1)} (x,m_G^2,m_G^2) =& (3 A(x) + 2 x \delta_{\ov{MS}}) \blog m_G^2 + \mathcal{O}(m_G^2)\\
f_{VVS}^{(0,0,1)} (y, z; m_G^2, m_G^2) =& -\big(3 P_{SS} (y,z) + \delta_{\ov{\rm MS}}\big)\blog m_G^2 - R_{VV} (y,z) + \mathcal{O}(m_G^2).
\end{align}

\section{Diagrams regulated by momentum}
\label{SEC:momentum}
When studying the mass terms, we encountered some diagrams for which the resummation of the Goldstone contributions provide no 
shift to regulate an infrared divergence and hence these diagrams must be regulated by momentum. More precisely, this is the case for the functions 
$U$, $M$, $Z$ and for some of the $V$ diagrams. 
In this section, we give the expansions for small external momentum $s\equiv-p^2$ of the diagrams 
that diverge as $s\rightarrow 0$, 
taken from expanding expressions in \cite{Martin:2003qz,Scharf:1993ds} or found by newly solving or expanding the  
integral equations in \cite{Martin:2003qz}. Hence we stress that (most of) this section contains new results not found elsewhere.

First, for $Z$, we only need the fact that 
\begin{equation}
 B(p^2,\gbm,\gbm)\underset{m_G\rightarrow 0}{\longrightarrow}2-\llog(-s).
\end{equation}

Then, for the $U$ function, taking one argument to zero does not cause any divergence, 
and we find, looking at the integral definition~(\ref{U_integral}) of $U$, that $U(x,y,0,0)$, $U(0,x,y,0)$, $U(x,0,y,0)$, $U(x,0,0,0)$ and $U(0,y,0,0)$ are all regular so we can substitute them for $U_0 + \mathcal{O}(s)$. The 
only divergent function is  $U(0,0,x,y)$ that has the form
\begin{align}
 U(0,0,x,y)=A_U(x,y)\llog(-s)+B_U(x,y)+\mathcal{O}(s)
\end{align}
with
\begin{align}
 A_U(x,y)=&-1+\frac{x \llog x - y \llog y}{x-y} = P_{SS} (x,y) \rightarrow A_U(x,x)= \llog x, \\
 B_U(x,y)=&\frac{5}{2}+ \frac{1}{2(y-x)}\bigg[ -(x+y) \llog^2 y +4x\llog x-4y\llog y\nn\\
 &\quad\quad\quad\quad\quad\quad\quad+2x\llog x\llog y- 2 (x+y) \mathrm{Li}_2 \left(1-\frac{x}{y}\right) \bigg]\nn\\
=& \frac{5}{2}+ \frac{1}{2(y-x)} \bigg[ 8 ( x \llog x - y \llog y) + (x+y) \big( \llog^2 x - \llog^2 y\big)  \nn\\
& - 2 (y-x) \llog x \llog y  - (x+y) \bigg(\mathrm{Li}_2 \bigg(1-\frac{x}{y}\bigg) - \mathrm{Li}_2 \bigg(1-\frac{y}{x}\bigg)\bigg) \bigg], \\
B_U(x,x) =& - \frac{3}{2} - 3 \llog x - \frac{1}{2} \llog^2 x,
\end{align}
where we have written the $B_U$ coefficient in two ways, one for computational simplicity, and the other to explicitly show the symmetry in $x \leftrightarrow y$. The limit as $x \rightarrow 0 $ can be smoothly taken to give
\begin{equation}
 U(0,0,0,u)=(\llog u-1)\llog(-s)-\frac{\pi^2}{6}+\frac{5}{2}-2\llog u-\frac{1}{2}\llog^2u+\mathcal{O}(s).
\end{equation}
which matches an expansion of the full momentum-dependence expression in equation (6.24) of \cite{Martin:2003qz}.

Finally, 
\begin{equation*}
U(0,0,0,0) = \frac{1}{2} \big( \llog (-s) - 3 \big)^2 + 1
\end{equation*}
 is not required, 
as it always appears with $\lambda^{GGG}$ as a factor, which is zero up to higher order corrections.

Turning now to the $M$ function, there are more cases to consider. In the case of only one argument vanishing, we see from the integral expression
~(\ref{M_integral}) that the function is regular. From eqs.(6.28) and (6.31) in \cite{Martin:2003qz}, we also 
find that $M(x,y,z,0,0)$, $M(x,y,0,0,v)$ and $M(x,y,0,0,0)$ are finite. 
Then we have 
\begin{equation}
\label{exprM0y0uv}
 M(0,y,0,u,v) = A_M(y,u,v) \llog(-s) + B_M(y,u,v)
\end{equation}
where
\begin{align}
 A_M(y,u,v)=&\frac{u\llog u}{(y-u)(u-v)}-\frac{y\llog y}{(y-u)(y-v)}-\frac{v\llog v}{(y-v)(u-v)},\\
 B_M(y,u,v) =& -(2 + \llog v) A_M(y,u,v) \nn\\
&  + \frac{u + v}{(y-u)(u-v)} \mathrm{Li}_2 (1 - u/v) - \frac{v + y}{(y-u)(y-v)} \mathrm{Li}_2( 1 - y/v) .
\end{align}

$M(0,0,0,u,v)$ (resp. $M(0,y,0,u,0)$) is found by taking the limit $y\rightarrow 0$ (resp. $v\rightarrow 0$), which is regular, 
in the expression of $M(0,y,0,u,v)$.

The expression of $M(x,0,0,0,0)$ with full momentum dependence can be found in equation (6.31) of \cite{Martin:2003qz}, and becomes when expanding to leading order for small $s$
\begin{align}
 M(x,0,0,0,0)=
\frac{1}{6x} \bigg( 18 + \pi^2 - 12 \log\left(-s/x\right) + 3 \log^2\left(-s/x\right) \bigg).
\end{align}
Finally we find $M(0,0,0,0,v)$ to be
\begin{align}
M(0,0,0,0,v)=\frac{1}{v} \bigg( \log^2 (-s/v) - 2 \log (-s/v) + \frac{\pi^2}{3}  \bigg).
\end{align}

The approximate formulae for $U(0,0,x,y)$, $M(0,y,0,u,v)$ and $M(0,0,0,0,v)$ have been checked against the numerical results from {\tt TSIL} \cite{Martin:2005qm} and show 
excellent agreement until $s$ becomes of the order of the arguments in the functions -- even when $s$ is of the order of the mass parameters, the difference 
between the approximate result and the numerical from {\tt TSIL} is about $10\%$.

\bibliographystyle{utphys}
\bibliography{GGBC}

\end{document}